\documentclass[prb,showpacs,preprintnumbers,amsfonts,amssymb,amsmath,floats,twocolumn,aps]{revtex4}

\usepackage{graphicx}
\usepackage{dcolumn}
\usepackage{bm}
\usepackage[final,dvips]{epsfig}
\usepackage{bm}
\usepackage{color}

\begin{document} 
  
\title[Self-energy-functional approach: Analytical results and the Mott-Hubbard transition]
      {Self-energy-functional approach: Analytical results and the Mott-Hubbard transition} 

\author{Michael Potthoff}
\email{potthoff@physik.uni-wuerzburg.de}

\affiliation{
Institut f\"ur Theoretische Physik und Astrophysik, 
Universit\"at W\"urzburg, 
Germany
}

\begin{abstract}
The self-energy-functional approach proposed recently is applied to 
the single-band Hubbard model at half-filling to study the Mott-Hubbard 
metal-insulator transition within the most simple but non-trivial approximation.
This leads to a mean-field approach which is interesting conceptually:
Trial self-energies from a two-site single-impurity Anderson model are 
used to evaluate an exact and general variational principle.
While this restriction of the domain of the functional 
represents a strong approximation, the approach 
is still thermodynamically consistent by construction and represents
a conceptual improvement of the ``linearized DMFT'' which has been
suggested previously as a handy approach to study the critical regime
close to the transition.
It turns out that the two-site approximation is able to reproduce the complete 
(zero and finite-temperature) phase diagram for the Mott transition.
For the critical point at $T=0$, the entire calculation can be done 
analytically.
This calculation elucidates different general aspects of the 
self-energy-functional theory.
Furthermore, it is shown how to deal with a number of technical 
difficulties which appear when the self-energy functional is evaluated
in practice.
\end{abstract} 
 

\maketitle 

\section{Introduction}

The correlation-driven transition from a paramagnetic metal to a 
paramagnetic insulator (Mott-Hubbard transition \cite{IFT98,Geb97,Mot90}) 
is one of the most interesting problems in condensed-matter physics. 
As a prime example for a quantum-phase transition, the Mott-Hubbard 
transition is important from the physical point of view but also for the
development and test of general theoretical methods to treat correlated 
electron systems.
The minimum model required to study the Mott-Hubbard transition is the 
single-band Hubbard model. \cite{Hub63,Gut63,Kan63}
Inherent to this model is the competition between the electrons' kinetic 
energy which tends to delocalize the electrons and favors a metallic state
and the on-site Coulomb interaction which tends to localize the electrons
to avoid double occupancies and thereby favors an insulating state at half
filling.
Except for the one-dimensional case, \cite{LW60} however, exact results
with regard to the nature of the transition and the critical interaction
strength $U_{\rm c}$ are not available -- even for this highly simplified 
model system.
A direct numerical solution using exact-diagonalization or quantum Monte-Carlo 
methods \cite{Dag94} suffers from the difficulty to access the thermodynamic 
limit or the low-temperature, low-energy regimes.

Considerable progress has been made in recent years due to the development
of the dynamical mean-field theory (DMFT) \cite{GK92a,Jar92,GKKR96} which
focuses on the opposite limit of infinite spatial dimensions $D=\infty$.
\cite{MV89,MH89a,BM89,BM90}
Within the DMFT the problem is simplified by mapping the original lattice 
model onto an impurity model the parameters of which must be determined by 
a self-consistency condition.
Different techniques to solve the effective impurity model have been 
employed to study the Mott transition within the DMFT, 
iterative perturbation theory, \cite{GK92a,GK93,RKZ94}
exact diagonalization, \cite{CK94,RMK94,SRKR94,EGK+03}
renormalization-group methods, \cite{MSK+95,Bul99,BCV01}
and quantum Monte-Carlo. \cite{HF86,SJvD+99,RCK99,JO01,Blu02}
One of the most important characteristic of the transition is the value of 
the critical interaction strength $U_{\rm c}$ at zero temperature.
Roughly, the different techniques to solve the mean-field equations
predict $U_{\rm c}/W \approx 1 - 1.5$ 
where $W$ is the width of the free density of states.

Recently, a self-energy-functional approach (SFA) has been put forward. 
\cite{Pot03a}
The SFA is a general variational approach to correlated lattice models
where the grand potential $\Omega$ is considered as a functional of the
self-energy ${\bm \Sigma}$.
As this functional is constructed from an infinite series of renormalized
skeleton diagrams, it is not known in an explicit form and the variational
principle $\delta \Omega[{\bm \Sigma}] = 0$ cannot be exploited {\em directly}.
Usually, one replaces the exact but unknown
functional with an explicitly known but approximate one -- this is 
essentially the standard diagrammatic approach \cite{LW60} which
leads to weak-coupling approximations in the end. \cite{BK61}
Opposed to this weak-coupling perturbational approach, the functional 
dependence $\Omega[{\bm \Sigma}]$ is not approximated at all in the SFA.
The key observation is that the functional, though unknown explicitly,
can be evaluated on a restricted domain of trial self-energies ${\cal S}$.
The evaluation of the functional is exact, while the approximation is 
due to the fact that the self-energy in the variational principle is no 
longer considered as arbitrary.
In this way, depending on the choice for the space ${\cal S}$, some 
well-known but also some novel approximations can be realized.
Here we are interested in the single-band Hubbard model with Hamiltonian
$H$. As argued in Ref.\ \onlinecite{Pot03a} a useful trial self-energy has
to be constructed as the exact self-energy of a different model 
(``reference system'') with Hamiltonian $H'$.
The reference system can be chosen arbitrarily -- it must, however,
share the same interaction part with the original model $H$.
The variational parameters at one's disposal are therefore the 
one-particle parameters of the reference system ${\bm t}'$.
The trial self-energy is parameterized as 
${\bm \Sigma} = {\bm \Sigma}({\bm t}')$, and the variational principle
reads $\partial \Omega[{\bm \Sigma}({\bm t}')] / \partial {\bm t}' = 0$.
To provide trial self-energies is the only purpose of the reference 
system $H'$. Whenever one is able to compute ${\bm \Sigma}$ for the
reference system $H'$, an exact evaluation of 
$\Omega[{\bm \Sigma}({\bm t}')]$ is possible.

Choosing $H'$ to be a system of decoupled sites, yields a Hubbard-I-type
approximation.
An improved approximation is obtained when $H'$ consists of decoupled 
clusters with a finite number of sites $N_{\rm c}>1$ per cluster 
as has been considered in Refs.\ \onlinecite{PAD03,DAH+03}.
This approach not only recovers the so-called cluster-perturbation 
theory (CPT) \cite{GV93,SPPL00,ZEAH02} but also gives a variational
improvement (V-CPT) which e.g.\ allows to describe phases with 
spontaneously broken symmetry. \cite{DAH+03}
Another possibility is to take $N_{\rm c}=1$, which implies the trial
self-energy to be local, but to include a coupling to a number of 
$n_{\rm b}$ additional uncorrelated (``bath'') sites.
In this case the reference system consists of a decoupled set of 
single-impurity 
Anderson models (SIAM) with $n_{\rm s} = 1 + n_{\rm b}$ sites each.
As has been shown in Ref.\ \onlinecite{Pot03a}, this approach not only
recovers the DMFT (namely in the limit $n_{\rm b} \to \infty$) but 
also provides a new variant of the exact-diagonalization 
approach, namely for any finite $n_{\rm b}$.
As compared to previous DMFT-exact-diagonalization approaches, 
\cite{CK94,RMK94,SRKR94,EGK+03} the construction
gives a thermodynamically consistent approximation even for small 
$n_{\rm b}$.
More complicated reference systems may be taken for the construction 
of consistent approximations, for example a system of decoupled clusters
of size $N_{\rm c}$ where each site in the cluster is coupled to 
$n_{\rm b}$ additional bath sites.
It has been shown \cite{PAD03} that in the limit $n_{\rm b} \to \infty$ 
the cellular DMFT (C-DMFT) is obtained, \cite{KSPB01} while approximations 
with finite $n_{\rm b}$ represent cluster approximations which ``interpolate'' 
between the CPT ($n_{\rm b} = 0$) and the C-DMFT ($n_{\rm b} = \infty$).

In the present paper the Mott transition is studied within the most simple
but non-trivial 
approximation: 
The variational principle $\delta \Omega[{\bm \Sigma}] = 0$ is exploited 
using a local trial self-energy from a reference system with $N_{\rm c}=1$ 
and a single additional bath site only, $n_{\rm b} = 1$. 
As $n_{\rm s}=1+n_{\rm b}=2$
this approximation will be referred to as the two-site dynamical impurity
approximation ($n_{\rm s}=2$-DIA) in the following.

The paper is organized as follows:
A brief review of the self-energy-functional approach will be given in the 
next section \ref{sec:sfa}.
The general aspects of the evaluation of the SFA are discussed in 
Sec.\ \ref{sec:eval} while Sec.\ \ref{sec:local} focuses on local
approximations ($N_{\rm c}=1$, $n_{\rm s}$ arbitrary) in particular.
In Sec.\ \ref{sec:twosite} the further specialization to the case 
$n_{\rm s} = 2$ is considered. 
The $n_{\rm s}=2$ dynamical-impurity approximation is motivated
(i) by  the fact that at the critical point for the Mott transition the entire 
calculation can be done {\em analytically}, 
(ii) by the conceptual simplicity of the approach which rests on a {\em single}
approximation only
and (iii) by making contact with a linearized DMFT (L-DMFT)
\cite{BP00,PN99d,OBH01,OBHP01,OPB03,Pot01} developed previously.
This is discussed in detail in Sec.\ \ref{sec:twosite}
while Sec.\ \ref{sec:lin} then presents the analytical calculation for the 
critical regime.
The results are discussed in Sec.\ \ref{sec:results}. 
The complete phase diagram for $T=0$ and finite temperatures is
addressed in Sec.\ \ref{sec:temp} and the
conclusions are given in Sec.\ \ref{sec:con}.

\section{Self-energy-functional theory}
\label{sec:sfa}

Useful information on correlated electron systems can be gained 
by exact-diagonalization 
or quantum Monte-Carlo methods applied to a lattice of finite size. \cite{Dag94}
This approach, however, suffers from the difficulty to access the thermodynamic 
limit and is therefore of limited use to describe phases with long-order and 
phase transitions.
On the other hand, using an {\em embedding approximation}, one can directly 
work in the thermodynamic limit and describe phase transitions while the 
actual numerical treatment has to be done a system of finite size only.
In the context of an embedding technique we have to distinguish between the 
original model of infinite size $H$ and an (e.g.\ spatially) truncated 
{\em reference system} $H'$.
The reference system must not necessarily be finite but it may consist of
an infinite number of decoupled subsystems with a finite number of degrees 
of freedom each.
In any case, $H'$ must be exactly solvable.

If one is interested not only in the equilibrium thermodynamics but also in the
elementary one-particle excitations, an embedding technique should focus on a 
dynamical quantity, such as the frequency-dependent self-energy ${\bm \Sigma}$.
The knowledge of ${\bm \Sigma}$ then allows to derive a thermodynamic potential 
$\Omega$ as well as different static and dynamic quantities via general relations.
The main steps are the following:
(i) Truncate the original model $H$ to obtain a simpler model $H'$
which is tractable numerically.
(ii) Calculate the self-energy ${\bm \Sigma}'$ of the reference system. 
(iii) Use ${\bm \Sigma} = {\bm \Sigma}'$ as an approximation for the
self-energy of $H$ and determine the grand potential $\Omega$ as well as 
further quantities of interest.
(iv) To get a self-consistent scheme, optimize the parameters of $H'$ by a 
feedback from the approximate solution at hand.
Ideally, the last step should be based on a general variational principle.
This is exactly the strategy of the self-energy-functional approach (SFA)
proposed recently. \cite{Pot03a}
A brief review of the essentials of this approach is given in the following.

Consider a system of fermions on an infinite lattice with on-site Coulomb
interaction at temperature $T$ and chemical potential $\mu$.
Its Hamiltonian $H = H_0({\bm t}) + H_1({\bm U})$ consists of a 
one-particle part which depends on a set of hopping parameters ${\bm t}$ 
and an interaction part with Coulomb-interaction parameters ${\bm U}$:
\begin{equation}
  H = \sum_{\alpha\beta} t_{\alpha\beta} 
  c_{\alpha}^\dagger c_{\beta}
  + \frac{1}{2} 
  \sum_{\alpha\beta\gamma\delta} U_{\alpha\beta\delta\gamma}
  c_{\alpha}^\dagger c^\dagger_{\beta} c_{\gamma} c_{\delta}  \: .
\label{eq:ham}
\end{equation}
The grand potential $\Omega$ can be obtained from the stationary point of
a self-energy functional
\begin{equation}
  \Omega_{\bm t}[{\bm \Sigma}] \equiv 
  {\rm Tr} \ln (- ({\bm G}_0^{-1} - {\bm \Sigma})^{-1}) + F[{\bm \Sigma}]
\label{eq:omega}
\end{equation}
as has been discussed in Ref.\ \onlinecite{Pot03a}.
Here the subscript ${\bm t}$ indicates the parametric dependence of the
functional on the hopping. This dependence is {\em exclusively} due to 
${\bm G}_0 = 1 / (\omega + \mu - {\bm t})$, the free Green's function
of $H$. Further, $F[{\bm \Sigma}]$ is the 
Legendre transform of the Luttinger-Ward functional $\Phi[{\bm G}]$.
As the latter is constructed as an infinite series of renormalized 
skeleton diagrams, \cite{LW60} the self-energy functional is not known 
explicitely.
Nevertheless, the {\em exact} evaluation of $\Omega_{\bm t}[{\bm \Sigma}]$ 
and the determination of the stationary point is possible \cite{Pot03a} on a 
{\em restricted} space ${\cal S}$ of trial self-energies 
${\bm \Sigma}({\bm t}') \in {\cal S}$.
Due to this restriction the procedure becomes an approximation.

Generally, the space $\cal S$ consists of ${\bm t}'$ representable self-energies.
${\bm \Sigma}$ is termed ${\bm t}'$ representable if there are hopping 
parameters ${\bm t}'$ such that ${\bm \Sigma} = {\bm \Sigma}({\bm t}')$ is the 
exact self-energy of the model $H' = H_0({\bm t}') + H_1({\bm U})$ (``reference 
system'').
Note that both the original system $H$ and the reference system $H'$ must share 
the same interaction part.
For any ${\bm \Sigma}$ parameterized as ${\bm \Sigma}({\bm t}')$ 
we then have: \cite{Pot03a}
\begin{eqnarray}
   \Omega_{\bm t}[{\bm \Sigma}({\bm t}')] &=& \Omega'({\bm t}')
   \nonumber \\
   &+& {\rm Tr} \ln (- ({\bm G}_0({\bm t})^{-1}  - {\bm \Sigma}({\bm t}'))^{-1})
   \nonumber \\
   &-& {\rm Tr} \ln (- ({\bm G}_0({\bm t}')^{-1} - {\bm \Sigma}({\bm t}'))^{-1}) \: ,
\label{eq:om}
\end{eqnarray}
where $\Omega'({\bm t}')$, ${{\bm G}'_0} \equiv {{\bm G}_0}({\bm t}') 
= 1 / (\omega + \mu - {\bm t}')$, and ${\bm \Sigma}({\bm t}')$ are the grand 
potential, the free Green's function and the self-energy of the reference system 
$H'$ while ${\bm G}_0$ is the free Green's function of $H$.
For a proper choice of ${\bm t}'$, 
a (numerically) exact computation of these quantities is possible.
Hence, the self-energy functional (\ref{eq:om}) can be evaluated exactly 
for this ${\bm \Sigma} = {\bm \Sigma}({\bm t}')$.
A certain approximation is characterized by a choice for $\cal S$.
As ${\bm \Sigma}$ is parameterized by ${\bm t}'$, this means to specify a 
space of variational parameters ${\bm t}'$.
Any choice will lead by construction to a non-perturbative approach which is 
thermodynamically consistent as an explicit expression for a thermodynamical
potential is provided. 
It turns out that a stationary point of the self-energy functional is a saddle 
point in general.
As in different standard variational methods, such as in the time-dependent 
density-functional approach, \cite{RG84} in the Green's-function approach 
\cite{BK61} and also in a recently considered variant, \cite{CK01} this implies
that there is no strict upper bound for the grand potential.
For a further discussion of the general concepts of the SFA see 
Ref.\ \onlinecite{Pot03a}.

\begin{figure}[t]
\includegraphics[width=55mm]{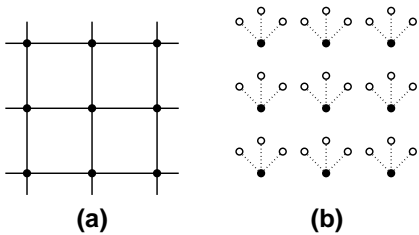}
\caption{
Schematic representation of the single-band Hubbard model $H$ (a)
and a possible reference system $H'$ (b).
$H'$ is a set of decoupled single-impurity Anderson models with
one correlated ($U>0$) impurity site and a number of $n_{\rm s}-1$
uncorrelated ($U=0$) bath sites each.
In the figure $n_{\rm s}=4$.
Note that the interaction part is the same for (a) and (b).
Variational parameters are the one-particle parameters of $H'$.
}
\label{fig:refsystem}
\end{figure}

So far the discussion is completely general. 
In Sec.\ \ref{sec:local} we will consider $H$ to 
be the Hubbard model and $H'$ to be a system of decoupled single-impurity 
Anderson models (SIAM).
Each SIAM consists of $n_{\rm s}$ sites, one correlated site (with $U>0$) and 
$n_{\rm s}-1$ uncorrelated ``bath'' sites ($U=0$).
This is illustrated by Fig.\ \ref{fig:refsystem} for $n_{\rm s}=4$.
Note that for any choice of $n_{\rm s}$, the original system and the 
reference system share the same interaction part -- as required by the
general theory.
The one-particle parameters of $H'$ are the variational parameters, i.e.\
the one-particle energies of the original sites and the bath sites and the
hopping (``hybridization'') between them.
As noted in Ref.\ \onlinecite{Pot03a}, the dynamical mean-field theory
(DMFT) is recovered in the limit $n_{\rm s} \to \infty$.
For $n_{\rm s} < \infty$ one obtains a new variant of the 
DMFT-exact-diagonalization approach. \cite{CK94,RMK94,SRKR94,EGK+03}
For a finite number of bath sites, the DMFT self-consistency condition
cannot be strictly satisfied. In the DMFT-ED method one therefore 
has to introduce a certain measure which allows to minimize the error
due to the discretization of the bath. The conceptual advantage of 
the SFA consists in the fact that this measure is replaced by a
variational procedure which is based on a physical variational principle.
As shown in Ref.\ \onlinecite{Pot03a}, a very good quantitative 
agreement with results from full DMFT calculations can be achieved
for the quasi-particle weight even with $n_{\rm s}=4$.
With $n_{\rm s} = 2$ a much simpler approach is considered here (Secs.\ 
\ref{sec:twosite} and \ref{sec:lin}) which, however, is still thermodynamically 
consistent and allows for simple and systematic investigations of 
the Mott transition.

\section{Evaluation of the self-energy functional}
\label{sec:eval}

The evaluation of the self-energy-functional theory can be done by solving the
Euler equation \cite{Pot03a} corresponding to the variational principle. 
While such an approach is possible in principle, it appears complicated as
dynamical two-particle quantities of the reference system are required.
An attractive alternative consists in the direct calculation of the grand 
potential along Eq.\ (\ref{eq:om}).
The numerical computation of $\Omega_{\bm t}[{\bm \Sigma}({\bm t}')]$
for a given set of one-particle parameters ${\bm t}'$ is straightforward for a 
reference system $H'$ of finite size.
There are, however, a few technical difficulties which appear in the practical
calculation and which shall be discussed in the following.
The problem of finding a stationary point of the function 
${\bm t}' \to \Omega_{\bm t}[{\bm \Sigma}({\bm t}')]$ is not addressed here
as this is a standard 
numerical problem very similar to the problem of finding a minimum of a real 
single-valued function of several arguments. 

$\Omega_{\bm t}[{\bm \Sigma}({\bm t}')]$ consists of three parts as given
by Eq.\ (\ref{eq:om}).
The grand potential of the reference system can be calculated as
$\Omega'({\bm t}') = - T \ln {\rm tr}' \exp(-(H' - \mu N')/T) =
- T \ln \sum_m \exp(-(E'_m - \mu N'_m)/T)$ from the many-body eigenenergies 
$E'_m - \mu N'_m$ of $H'-\mu N'$ where $N'$ is the total particle number operator.
Direct numerical diagonalization or (at $T=0$) the Lanczos technique \cite{LG93} 
may be used.

Next, the second term on the r.h.s.\ of Eq.\ (\ref{eq:om}) is discussed.
In the following the dependence of ${\bm \Sigma}$ on ${\bm t}'$ will be 
suppressed for convenience and its dependence on $\omega$ is made explicit
in the notations.
The diagonalization of $H' - \mu N'$ yields (via the Lehmann representation) 
the Green's function ${\bm G}'(\omega)$ and the free Green's function 
${\bm G}_0'(\omega)$ of the 
reference system. 
The self-energy 
is then obtained from the Dyson equation 
of the reference system
${\bm \Sigma}(\omega) = {{\bm G}_0'}^{-1}(\omega) - {{\bm G}'}^{-1}(\omega)$. 
Using the self-energy ${\bm \Sigma}(\omega)$, one obtains the (approximate) 
Green's function of the original model via
${\bm G}(\omega) \equiv ({\bm G}_0^{-1}(\omega) - {\bm \Sigma}(\omega))^{-1}$. 
The remaining task is to calculate ${\rm Tr} \ln (- {\bm G})$ for a lattice 
consisting of a finite number of sites $L$. The thermodynamic limit 
$L \to \infty$ is
performed in the end. Translational symmetry is not necessarily required.

It is important to note that ${\bm G}(\omega)$ 
is causal, i.e.\ ${\bm G}(\omega + i0^+) = 
{\bm G}_{\rm R}(\omega) - i {\bm G}_{\rm I}(\omega)$ with 
${\bm G}_{\rm R}(\omega)$, ${\bm G}_{\rm I}(\omega)$ Hermitian and 
${\bm G}_{\rm I}(\omega)$ 
positive definite for any real $\omega$
($0^+$ is a positive infinitesimal).
The causality of the Green's function ${\bm G}(\omega)$ is ensured by the 
causality of ${\bm \Sigma}(\omega)$ and ${\bm G}_0(\omega)$, see
Appendix \ref{sec:causal}.
The latter are causal as these are exact quantities.

Let $\omega_m$ be the (real, first-order) poles of ${\bm G}(\omega)$. 
For $\omega \to \omega_m$ we have 
${\bm G}(\omega) \to {\bm R}_m / (\omega - \omega_m)$ where due to the causality 
of ${\bm G}(\omega)$ the matrix ${\bm R}_m$ is positive definite. 
Therefore, the frequency-dependent diagonalization of 
${\bm G}(\omega) = {\bm U}(\omega) {\bm g}(\omega) {\bm U}(\omega)^\dagger$
with unitary ${\bm U}(\omega)$ (for real $\omega$) 
yields a diagonal Green's function ${\bm g}(\omega)$
with elements $g_k(\omega)$ that are real for real $\omega$ and have first-order 
poles at $\omega = \omega_m$ with {\em positive} residues.

As the Green's function can be written as 
${\bm G}(\omega) = 1 / (\omega + \mu - {\bm t} - {\bm \Sigma}(\omega))$, 
the unitary transformation ${\bm U}(\omega)$ also diagonalizes the real and
symmetric matrix ${\bm t} + {\bm \Sigma}(\omega)$, i.e.\ 
${\bm t} + {\bm \Sigma}(\omega) = 
{\bm U}(\omega) {\bm \eta}(\omega) {\bm U}(\omega)^\dagger$
and $g_k(\omega) = 1 / (\omega + \mu - \eta_k(\omega))$.
The $\eta_k(\omega)$ are real for real $\omega$ and have first-order 
poles at $\omega = \zeta_n$ with $\zeta_n$ being the poles of the self-energy.
For $\omega \to \zeta_n$ we have 
${\bm \Sigma}(\omega) \to {\bm S}_n / (\omega - \zeta_n)$ with positive definite
${\bm S}_n$. 
Consequently, the residues of $\eta_k(\omega)$ at $\omega = \zeta_n$ are {\em positive}.

One can write  
$g_k(\omega) = \sum_m R_{k,m} / (\omega - \omega_m) + \widetilde{g}_k(\omega)$
and 
$\eta_k(\omega) = \sum_n S_{k,n} / (\omega - \zeta_n) + \widetilde{\eta}_k(\omega)$
with $R_{k,m}, S_{k,n} > 0$ and where 
$\widetilde{g}_k(\omega)$ and $\widetilde{\eta}_k(\omega)$ are analytical in
the entire $\omega$ plane. 
As ${\bm G}(\omega) \sim 1 / \omega$ and ${\bm \Sigma}(\omega) \sim \mbox{const.}$
for $\omega \to \infty$, one has $\widetilde{g}_k(\omega) = 0$ and 
$\widetilde{\eta}_k(\omega) = \mbox{const.}$ and real.
Consequently, $-(1/\pi)\mbox{Im}\, g_k(\omega + i 0^+) \ge 0$ and 
$-(1/\pi)\mbox{Im} \, \eta_k(\omega + i 0^+) \ge 0$.
This will be used in the following.

The trace ``Tr'' in Eq.\ (\ref{eq:om}) consists of a sum $T\sum_\omega$ 
over the fermionic Matsubara 
frequencies $i\omega = i(2n+1) \pi T$ ($n$ integer) and a trace ``tr'' with respect 
to the quantum numbers $\alpha$; see Eq.\ (\ref{eq:ham}).
The convergence of the frequency sum is ensured by the usual factor 
$\exp(i\omega 0^+)$ from the diagram rules.
The calculation then proceeds as follows:
\begin{eqnarray}
&&  
  T \sum_\omega e^{i\omega 0^+} {\rm tr} \: \ln \: 
  \frac{-1}{i\omega + \mu - {\bm t} - {\bm \Sigma}(i\omega)}
\nonumber \\ && \stackrel{\rm (a)}{=} 
  \frac{-1}{2\pi i} \sum_{k} \oint_C d\omega \: e^{\omega 0^+} f(\omega) \:
  \ln \: (-g_k(\omega) )
\nonumber \\ && \stackrel{\rm (b)}{=} 
  \frac{-1}{\pi} \sum_{k} \int_{-\infty}^\infty d\omega \: f(\omega) \:
  {\rm Im} \: \ln (-g_k(\omega + i0^+) )
\nonumber \\ && \stackrel{\rm (c)}{=} 
  - \sum_{k} \int_{-\infty}^\infty d\omega \: f(\omega) \:
  \Theta \left( \omega + \mu - \eta_k(\omega) \right)
\nonumber \\ && \stackrel{\rm (d)}{=} 
  - \sum_{k} \int_{-\infty}^\infty \!\! d \omega \:
  T \: \ln(1+e^{-\omega/T}) \: \frac{d
  \Theta(\omega + \mu - \eta_k(\omega))
  }{d\omega} 
\nonumber \\ && \stackrel{\rm (e)}{=}   
  - 2L \sum_{m} T \: \ln(1+\exp(-\omega_{m}/T)) 
  - R_\Sigma 
\label{eq:trlng}
\end{eqnarray}
with 
\begin{equation}
  R_\Sigma = - \sum_{n} T \: \ln(1+\exp(-\zeta_n/T)) \: .
\end{equation}
In Eq.\ (\ref{eq:trlng})
$\ln$ denotes the principal branch of the logarithm, $f(\omega)=1/(\exp(\omega/T)+1)$
is the Fermi function, and $C$ is a contour in the complex $\omega$ plane enclosing 
the first-order poles of the Fermi function in counterclockwise direction.
In step (a) the transformation ${\bm U}(\omega)$ is performed under the
trace.
Convergence of the integral for $\omega \to \pm \infty$ is ensured by the
Fermi function and by the factor $e^{\omega 0^+}$, respectively.
Step (b) results from analytical continuation to real frequencies.
In step (c) $-(1/\pi)\mbox{Im}\, g_k(\omega + i 0^+) \ge 0$ has been used
(see Appendix \ref{sec:fom}).
At this point the causality of the Green's function is essential, as discussed
above.
In step (d) the Fermi function is written as a derivative with respect to 
$\omega$ and integration by parts is performed.
Step (e) uses the results of Appendix \ref{sec:theta} for the derivative of the 
step function.
As the different diagonal elements of the Green's function, $g_k(\omega)$, have
the same set of poles and zeros,
\footnote{
  We formally allow for poles of $g_k(\omega)$ and $\eta_k(\omega)$ 
  with vanishing residue. Note that adding a pole with zero weight to 
  $g_k(\omega)$ can only be accomplished if simultaneously a pole with 
  zero weight is added to $\eta_k(\omega)$, and $\mbox{Tr} \ln (- {\bm G})$ 
  is unchanged.
}
the sum over $k$ becomes trivial and yields a
factor $2L$ only where $L$ is the dimension of the hopping matrix, i.e.\ the 
number of orbitals. 
The factor 2 accounts for the two spin directions.
The contribution from the poles of the self-energy (Appendix \ref{sec:theta}) 
is denoted by $R_\Sigma$.
Apart from this correction term, 
${\rm Tr} \ln (- ({\bm G}_0({\bm t})^{-1} - {\bm \Sigma}({\bm t}'))^{-1})$
turns out to be the grand potential of a system of {\em non-interacting} 
quasi-particles with {\em unit} weight and energies given by the poles of 
${\bm G}(\omega) \equiv ({\bm G}_0^{-1}(\omega) - {\bm \Sigma}(\omega))^{-1}$.

Consider now the third term on the r.h.s.\ of Eq.\ (\ref{eq:om}).
A calculation completely analogous to Eq.\ (\ref{eq:trlng}) results in:
\begin{eqnarray}
  && T \sum_\omega e^{i\omega 0^+} {\rm tr}' \: \ln (-{\bm G}'(i\omega)) 
\nonumber \\
  && = - 2L \sum_{m} T \: \ln (1+\exp(-{\omega'_m}/T))
   - R_\Sigma \: .
\label{eq:trlngs}
\end{eqnarray}
Here $\omega'_m$ are the poles of ${\bm G}'$.
Again, the first term in Eq.\ (\ref{eq:trlngs}) is the grand potential of a 
non-interacting system of fermions with one-particle energies given by 
the poles of the Green's function ${\bm G}'$.
The same holds for the second term, but with energies given by the poles of 
the self-energy.
By construction, the self-energy is the same for both, the original system 
and the reference system. 
Hence, the same correction term $R_\Sigma$ appears in Eqs.\ (\ref{eq:trlng}) 
and (\ref{eq:trlngs}) 
and cancels out in 
Eq.\ (\ref{eq:om}). 

Note that a pole of ${\bm G}(\omega)$ at $\omega = \omega_m$ with residue
${\bm R}_m \to 0$ implies a pole of ${\bm G}'(\omega)$ at the same frequency
$\omega_m$ (with residue ${\bm R}'_m \to 0$).
Hence, contributions due to poles with vanishing residues cancel out in 
Eq.\ (\ref{eq:om}).
The reason is the following:
Suppose that $g_k(\omega) = R_{k,m} / (\omega - \omega_m)$ for 
$\omega$ close to $\omega_m$ with residue $R_{k,m} \to 0$.
For the diagonal elements of ${\bm t} + {\bm \Sigma}(\omega)$ this implies
that $\eta_k(\omega) = (1/R_{k,m}) (\omega - \omega_m)$ near $\omega_m$.
A zero of $\eta_k(\omega)$ with infinite positive coefficient $1/R_{k,m}$
must be due to ${\bm \Sigma}(\omega)$. 
Therefore, for the diagonal elements of ${\bm t}' + {\bm \Sigma}(\omega)$ 
this means that $\eta'_k(\omega) = (1/R_{k,m}) (\omega - \omega_m)$.
Consequently, $g'_k(\omega) \equiv 1 / (\omega + \mu - \eta'_k(\omega))
= R_{k,m} / (\omega - \omega_m)$ for $\omega$ close to $\omega_m$ 
with $R_{k,m} \to 0$.
The argument also shows that although the residues do not appear in Eqs.\ 
(\ref{eq:trlng}) and (\ref{eq:trlngs}) explicitly, one can state that poles 
with small residues will give a small contribution in Eq.\ (\ref{eq:om}).

Based on the causality of the respective Green's functions, an efficient
algorithm can be set up to find $\omega_m$ and $\omega'_m$ numerically 
which are then needed in Eqs.\ (\ref{eq:trlng}) and (\ref{eq:trlngs}).
The poles of ${\bm G}'(\omega)$ are directly obtained from the diagonalization 
of the reference system.
The problem consists in finding the poles of ${\bm G}(\omega)$.
Allowing for poles with vanishing (very small) residue, one can assume the 
function $g'_k(\omega)$ for fixed but arbitrary $k$ to display all the poles 
of ${\bm G}'(\omega)$.
Since $g'_k(\omega)$ is of the form 
$g'_k(\omega) = \sum_m R'_{k,m} / (\omega - \omega'_m)$
with $R'_{k,m} \ge 0$, it is monotonically decreasing.
Hence, there is exactly one zero of $g'_k(\omega)$ located in the interval 
between two adjacent poles $\omega'_m$ and $\omega'_{m+1}$.
As $g'_k(\omega)$ is monotonous, the zero $\zeta_m$ can easily be found 
numerically by an iterative bisection procedure.
The zeros of $g'_k(\omega)$ are the poles of $\eta'_k(\omega)$ and the poles
of $\eta'_k(\omega)$ are the same as the poles of $\eta_k(\omega)$.
Now, since $g_k(\omega) = 1 / (\omega + \mu - \eta_k(\omega))$, 
$g_k(\omega)$ and $g'_k(\omega)$ must have the same set of zeros.
The function $g_k(\omega)$ is monotonically decreasing.
Therefore, the poles of $g_k(\omega)$ can be found between the $\zeta_m$
by using the same iterative bisection procedure once again.

\section{Local approximations}
\label{sec:local}

In the following we consider $H$ to be the single-band Hubbard model:
\begin{equation}
  H = \sum_{ij\sigma} t_{ij} c_{i\sigma}^\dagger c_{j\sigma} 
  + \frac{U}{2} \sum_{i\sigma} n_{i\sigma} n_{i-\sigma} \: .
\end{equation}
For the reference system $H'$ shown in Fig.\ \ref{fig:refsystem}, the 
self-energy is local: $\Sigma_{ij}(\omega) = \delta_{ij} \Sigma(\omega)$.
Clearly, the approximation is the better the more degrees of freedom are
included in $H'$. 
The optimal local approximation is obtained with the most flexible (but local)
trial self-energy. 
This is the DMFT which is recovered for an infinite number of uncorrelated 
bath sites (per original correlated site), i.e.\ for $n_{\rm s}-1 \to \infty$.
On the other hand, $n_{\rm s}=1$ corresponds to a Hubbard-I-type approximation.
Here it will be shown that, for arbitrary $n_{\rm s}$, the evaluation of the
self-energy functional reduces to a one-dimensional integration only.
This is an important simplification for any practical numerical (or even
analytical) calculations.

In case that the self-energy is local it is advantageous to start from
step (c) in Eq.\ (\ref{eq:trlng}).
Assuming translational symmetry, 
the matrix ${\bm t} + {\bm \Sigma}(\omega)$ is diagonalized by Fourier
transformation to reciprocal space. 
Its eigenvalues $\eta_k(\omega)$ are given by
$\eta_k(\omega) = \epsilon({\bm k}) + \Sigma(\omega)$ where
$k=({\bm k},\sigma)$ and $\epsilon({\bm k})$ is the tight-binding Bloch dispersion. 
The self-energy is taken to be spin-independent and independent of the site 
index, i.e.\ a paramagnetic homogeneous phase is assumed for simplicity.
The second term on the r.h.s.\ of Eq.\ (\ref{eq:om}) then becomes:
\begin{eqnarray}
&&  
  {\rm Tr} \ln (- ({\bm G}_0({\bm t})^{-1} - {\bm \Sigma}({\bm t}'))^{-1})
\nonumber \\ && \stackrel{\rm (c)}{=} 
  - \sum_{{\bm k}\sigma} \int d\omega \: f(\omega) \:
  \Theta \left( \omega + \mu - \epsilon({\bm k}) - \Sigma(\omega) \right)
\nonumber \\ && = 
  - 2L \int d\omega \: f(\omega)  \int dz \, 
  \rho_0(z) \, \Theta \left( \omega + \mu - z - \Sigma(\omega) \right)
\nonumber \\ && =
  2L \int d\omega \: f(\omega)  \int dz \, 
  R_0(z) \, \frac{d}{dz} \Theta \left( \omega + \mu - z - \Sigma(\omega) \right)
\nonumber \\ && =
  - 2L \int d\omega \: f(\omega) \: 
  R_0(\omega + \mu - \Sigma(\omega)) \: .
\label{eq:trlng1}
\end{eqnarray}
Here $L$ is the number of lattice sites, the factor 2 stems for the spin summation,
$\rho_0(z) = L^{-1} \sum_{{\bm k}} \delta(z-\epsilon({\bm k}))$ is the non-interacting
density of states, and $R_0(z)=\int_{-\infty}^z dz' \rho_0(z')$ is an antiderivative.
Note that there is no ``correction term'' $R_\Sigma$, as the derivative in 
Eq.\ (\ref{eq:trlng1}) is with respect to $z$.

The reference system $H'$ is a set of decoupled single-impurity Anderson models
with one correlated (``impurity'') site and $n_{\rm s}-1$ bath sites each.
The Hamiltonian is $H'= \sum_i H'(i)$ with
\begin{eqnarray}
  H'(i) &=& \sum_{\sigma} \epsilon_1 c_{i\sigma}^\dagger c_{i\sigma} 
            + \frac{U}{2} \sum_{\sigma} n_{i\sigma} n_{i-\sigma} 
	    \nonumber \\
	&+& \sum_{\sigma,k=2}^{n_{\rm s}} \epsilon_k a^\dagger_{ik\sigma} a_{ik\sigma}
	 +  \sum_{\sigma,k} \left( V_k c_{i\sigma}^\dagger a_{ik\sigma}  + \mbox{H.c.} 
	 \right) \: .
	 \nonumber \\
\label{eq:siam2}
\end{eqnarray}
The parameters $\epsilon_1$, $\epsilon_k$, and $V_k$ for $k=2,...,n_{\rm s}$ are
the on-site energies of the impurity site and of the bath sites, and the 
hybridization between them, respectively.
These are the variational parameters of the theory.

The third term on the r.h.s.\ of Eq.\ (\ref{eq:om}) reads:
\begin{eqnarray}
  && 
  {\rm Tr} \ln (- ({{\bm G}'_0}({\bm t}')^{-1} - {\bm \Sigma}({\bm t}'))^{-1})
\nonumber \\
  && = 2LT \sum_\omega e^{i\omega 0^+} \left(
  \ln (-G'_1(i\omega))
  + \sum_{k=2}^{n_{\rm s}} \ln (-G'_k(i\omega))
  \right)
\nonumber \\
  && = -2L \int_{-\infty}^\infty d\omega \: f(\omega)
  \left(
  \Theta(G'_1(\omega)) + \sum_{k=2}^{n_{\rm s}} \Theta(G'_k(\omega))
  \right)
\nonumber \\
\label{eq:trlngs1}
\end{eqnarray}
where the first equation is derived in Appendix \ref{sec:hur} and
\begin{equation}
  G'_1(\omega) = \frac{1}{\omega + \mu - \epsilon_1 - \Delta(\omega) - \Sigma(\omega)}
\label{eq:g1def}
\end{equation}
is the impurity Green's function,
\begin{equation}
  G'_k(\omega) = \frac{1}{\omega + \mu - \epsilon_k}
\label{eq:grdef}
\end{equation}
is the (free) conduction-band Green's function, and
\begin{equation}
  \Delta(\omega) = \sum_{k=2}^{n_{\rm s}} \frac{V_k^2}{\omega + \mu - \epsilon_k}
\end{equation}
is the hybridization function.
The final expressions (\ref{eq:trlng1}) and (\ref{eq:trlngs1}) involve 
one-dimensional integrations only and can therefore be calculated numerically
without serious problems.

\section{Two-site dynamical impurity approximation}
\label{sec:twosite}

In the following we will focus on the case $n_{\rm s}=2$, i.e.\ on the two-site
dynamical-impurity approximation ($n_{\rm s}=2$-DIA).
There are different intentions which are followed up:

(i) For any approximation within the context of the SFA, one has to 
compute the self-energy of the reference system $H'$.
As the interaction part is the same for both, $H$ and $H'$, this still 
constitutes a non-trivial many-body problem which can only be treated
by numerical means in most cases.
The reference system characterized by $N_{\rm c}=1$ and $n_{\rm s} = 2$,
represents an exception:
For a special point in the space of model parameters (zero temperature, 
half-filling and $U=U_{\rm c}$, the critical interaction for the Mott transition), 
the entire calculation can be done {\em analytically}.
This not only includes the diagonalization of the reference system which 
actually is a simple dimer model but also and more important here, the 
exact evaluation of the self-energy functional for the trial self-energies 
considered and the subsequent variational optimization.
Therefore, the study of the $n_{\rm s}=2$-DIA is ideally suited to elucidate 
different technical points which are relevant for any $N_{\rm c}$ and $n_{\rm s}$
and which must be considered carefully.

(ii) The $n_{\rm s}=2$-DIA must be considered as {\em inferior} when compared 
to approximations with higher $n_{\rm s}$ and when compared to
$n_{\rm s} = \infty$ (the DMFT) in particular.
On the other hand, one has to keep in mind that the DMFT must always be
supplemented by an additional (numerical) method to solve the mean-field
equations which necessarily involves additional approximations.
Even if the additional approximations can be controlled, DMFT results 
always depend on the accuracy of the numerical method employed.
As concerns the $n_{\rm s}=2$-DIA, there is no such difficulty:
The theory rests on a {\em single} approximation only, namely the restriction 
of the space ${\cal S}$ to self-energies representable by the two-site 
reference system -- the rest of the calculation is rigorous.
It is this conceptual simplicity which makes the approximation
attractive.

(iii) That an approach referring to an $n_{\rm s}=2$-site SIAM is able 
to give reasonable results has been shown beforehand by the linearized
DMFT (L-DMFT). \cite{BP00,PN99d,OBH01,OBHP01,OPB03}
The L-DMFT is a well-motivated but ad-hoc simplification of the 
full DMFT and maps the Hubbard model self-consistently onto the
$n_{\rm s}=2$-site SIAM just at the critical point for the Mott 
transition.
The L-DMFT can be also be considered to represent the lowest-order 
realization of a more general projective self-consistent method (PSCM).
\cite{MSK+95}
As compared to the full DMFT, the linearized theory yields surprisingly
good estimates for the critical interaction $U_{\rm c}$ in the single-band
model, on translation invariant lattices \cite{BP00} as well as 
for lattice geometries with reduced translational symmetry. \cite{PN99d}
The approach can also be extended beyond the critical regime. \cite{Pot01}
The main disadvantage of the L-DMFT is that it not 
consistently derived from a thermodynamical potential.
Another intention of the present paper is therefore to suggest a two-site
method that conceptually improves upon the L-DMFT in this respect.
In fact, as the parameters of the effective $n_{\rm s}=2$-site impurity 
model are determined via a physically meaningful variational principle, the
two-site approximation within the SFA should be regarded as an {\em optimal}
two-site approach.
The interesting question is, of course, whether or not this improvement 
of the method also implies improved results.
To this end the results from the analytical evaluation of the $n_{\rm s}=2$-DIA
have to be compared with those of the L-DMFT and with available numerical 
results for $n_{\rm s}=\infty$ (full DMFT).
This represents a good check of the practicability of the new method.
As the $n_{\rm s}=2$-site DIA still represents a very handy method, it can
also be employed to investigate overall trends. 
It will be interesting to study the critical interaction $U_{\rm c}$ 
for a variety of different geometries (i.e.\ for different free densities 
of states).

\begin{figure}[t]
\includegraphics[width=75mm]{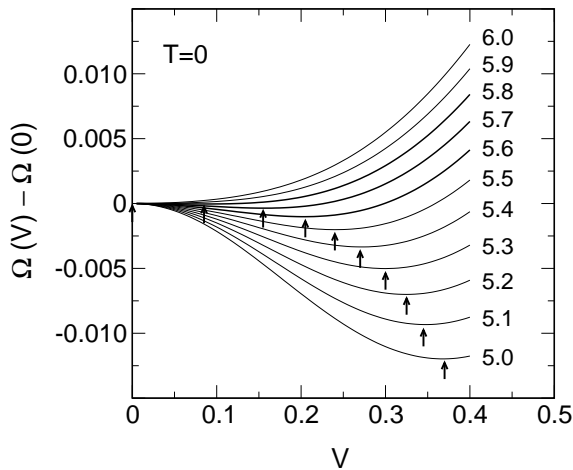}
\caption{
The grand potential $\Omega$ (per site) as calculated from 
Eq.\ (\ref{eq:om}) and Eqs.\ (\ref{eq:trlng1}) and (\ref{eq:trlngs1}) 
for the reference system with $n_{\rm s}=2$ as a function of the 
hybridization strength $V$ (only the difference $\Omega(V) - \Omega(V=0)$ 
is plotted).
Calculations for the Hubbard model at zero temperature $T=0$ and half-filling 
($\mu=U/2$). The free density of states is taken to be semi-elliptical with a
band width $W=4$. The interaction strength is varied from $U=5$ to $U=6$ as 
indicated. 
Arrows indicate stationary points of the self-energy functional.
The Mott transition takes place at a critical interaction strength 
$U_{\rm c} \approx 5.85$.
}
\label{fig:omega}
\end{figure}

Let us first consider a numerical evaluation of the theory.
Fig.\ \ref{fig:omega} shows the results of a numerical calculation along
the lines discussed above for the paramagnetic phase of the Hubbard model
at half-filling and zero temperature.
The free density of states (DOS) $\rho_0(z)$ is taken to be semi-elliptical with
a band width $W=4$.
The calculations are performed using the $n_{\rm s}=2$-DIA,
i.e.\ there is one bath site (per correlated site) only.
Due to manifest particle-hole symmetry, two of the variational parameters, the
on-site energies, are already fixed: $\epsilon_1=0$ and $\epsilon_2=\mu=U/2$.
As a function of the remaining variational parameter, the hybridization
strength $V\equiv V_{k=2}$, the grand potential $\Omega(V) = \Omega[{\bm \Sigma}(V)]$
shows two (non-equivalent) stationary points for $U=5$ (see Fig.\ \ref{fig:omega}):
a minimum at a finite $V=0.37$ and a maximum at $V=0$ (as $\Omega(V)=\Omega(-V)$, 
there is another minimum at $V=-0.37$ which can be ignored here).
The two stationary points correspond to two physically different phases:
For $V>0$ the interacting local density of states is finite at $\omega=0$ while
it vanishes for $V=0$. 
So there is a metallic and an insulating phase coexisting.
Due to the lower $\Omega$ at the respective stationary point, the metallic phase 
is stable as compared to the insulating one.
With increasing $U$ the optimal $V_{\rm met}$ and the energy difference 
$|\Omega(V_{\rm met}) - \Omega(0)|$ decrease.
For $U=U_{\rm c} \approx 5.85$ there is a metal-insulator (Mott-Hubbard) transition
which is characterized by a coalescence of the stable metallic with the metastable
insulating phase. For $U>U_{\rm c}$ there is the insulating phase only.

Qualitatively, this continuous transition is completely consistent with the
preformed-gap scenario \cite{MSK+95,GKKR96}
(however, see also Refs.\ \onlinecite{Keh98,NG99,Kot99}).
For $U < U_{\rm c}$ the self-energy is a two-pole function.
This leads to a three-peak structure in the interacting local density of states:
Similar as in the full DMFT, there are two Hubbard ``bands'' separated by 
an energy of the order of $U$, and a quasi-particle resonance at $\omega=0$.
On approaching the critical interaction $U_{\rm c}$ from below, the
weight $z$ of the resonance vanishes linearly $z \sim (U_{\rm c}-U)$ 
leaving a finite gap for $U>U_{\rm c}$.
As shown in Ref.\ \onlinecite{Pot03a} the quasi-particle weight calculated from
the self-energy at the respective optimal $V=V_{\rm met}$ is in a very good 
quantitative agreement with results from full DMFT calculations in the whole
range from $U=0$ to $U=U_{\rm c}$ -- for $n_{\rm s}=4$ and even for $n_{\rm s}=2$ 
which is the case considered here.

\section{``Linearized'' dynamical impurity approximation}
\label{sec:lin}

In the following we will concentrate on the critical regime $U \to U_{\rm c}$. 
It will be shown that the critical interaction strength can be calculated 
analytically for $n_{\rm s}=2$.
The independent analytical result can be compared with the numerical one of the 
preceeding section.
This represents a strong test of the numerics.

Consider the function $\Omega(V) = \Omega[{\bm \Sigma}(V)]$. 
As $\Omega(V)=\Omega(-V)$ there must be a stationary point of $\Omega(V)$ at 
$V=0$ for any $U$.
This implies that the linear term in an expansion around $V=0$ is missing, i.e.:
\begin{equation}
   \Omega(V) = \Omega(0) + A \cdot V^2 + {\cal O}(V^4) \: .
\end{equation}
The coefficient $A$ depends on $U$. 
Assuming that the metal is stable against the insulator for  $U < U_{\rm c}$, 
we must have $A < 0$ for $U < U_{\rm c}$ and $A > 0$ 
for $U > U_{\rm c}$.
This is a necessary condition for a continuous (``second-order'') transition
and consistent with the numerical results displayed in Fig.\ \ref{fig:omega}.
Therefore, the critical point for the Mott transition within the two-site
model is characterized by 
\begin{equation}
  A = 0 \: .
\label{eq:cond}
\end{equation}
The task is to calculate the three contributions to the grand potential 
following Eq.\ (\ref{eq:om}), to expand in $V$ up the the second-order term
and to find the interaction strength satisfying the condition (\ref{eq:cond}).

Consider the grand potential of the reference system first. 
With $\epsilon_1 = 0$, $\epsilon_2=U/2$ and $\mu=U/2$ the ground state of the 
two-site system, Eq.\ (\ref{eq:siam2}), lies is the invariant subspace with 
total particle number $N'=2$.
The ground-state energy $E'_0$ is readily calculated:
\begin{equation}
   E'_0 = \frac{3}{4} U - \frac{1}{4} \sqrt{U^2 + 64 V^2} \: .
\end{equation}
At $T=0$ the grand potential of the reference system is
$\Omega' = L E'_0 - L \mu \langle N' \rangle$. 
Therefore,
\begin{equation}
   \Omega'/L = - \frac{U}{4} - \frac{1}{4} \sqrt{U^2 + 64 V^2}
   = - \frac{U}{2} - \frac{8 V^2 }{ U} + {\cal O}(V^4) \: .
\label{eq:ucie}
\end{equation}
Actually, this is the grand potential {\em per site}.

We proceed with the third term on the r.h.s.\ of Eq.\ (\ref{eq:om}).
For the analytical calculation, it is convenient to start from 
Eq.\ (\ref{eq:trlngs}):
\begin{equation}
   T \sum_\omega {\rm tr}' \: \ln (-{\bm G}'(i\omega)) =
   2 L \sum_r \omega'_r \Theta(-\omega'_r) - R_\Sigma 
\end{equation}
where it has been used that $-T \: \ln (1+\exp(-\omega/T)) = \omega \Theta(-\omega)$
for $T=0$. The factor 2 is due to the spin degeneracy.
$R_\Sigma$ will cancel out later.
The Green's function of the two-site model is easily calculated.
There are four excitation energies, labeled by $r$, 
given by the four poles of the impurity 
Green's function at:
\begin{equation}
   \omega'_r = \pm \frac{1}{4} \left( 
   \sqrt{U^2 + 64 V^2} \pm \sqrt{U^2 + 16 V^2}
   \right) \: .
\end{equation}
This yields:
\begin{equation}
   T \sum_\omega {\rm tr}' \: \ln (-{\bm G}'(i\omega)) / L = -
   U - \frac{32V^2}{U} + {\cal O}(V^4) - R_\Sigma/ L \: .
\label{eq:ucimp}
\end{equation}

Finally, for the second term on the r.h.s.\ of Eq.\ (\ref{eq:om}) we have:
\begin{eqnarray}
   && T \sum_\omega {\rm tr} \: \ln (-{\bm G}(i\omega)) =
\nonumber \\ &&
   2 L \sum_r \int_{-\infty}^\infty dz \: \rho_0(z) \: 
   \omega_r(z) \Theta(-\omega_r(z)) 
   - R_\Sigma \: .
\label{eq:uclatt}
\end{eqnarray}
Here, Eq.\ (\ref{eq:trlng}) has been used with $k=({\bm k},\sigma)$, and the 
$\bm k$ sum has been replaced by an integration over $z$ weighted by the free 
DOS $\rho_0(z)$.
For a given $z=\epsilon({\bm k})$, the quasi-particle energies are obtained as
the poles of the lattice Green's function in reciprocal space, i.e.\ from the
solutions $\omega=\omega_r(z)$ of the equation $\omega + \mu - z - \Sigma(\omega) = 0$.
To find the roots, the self-energy of the reference system is needed:
\begin{equation}
   \Sigma(\omega) = \frac{U}{2} + \frac{U^2}{8} \left( 
   \frac{1}{\omega-3V} + \frac{1}{\omega+3V}
   \right) \: .
\end{equation}
This leads to a cubic equation:
\begin{equation}
  \omega^3 - z \omega^2 - (9 V^2 + U^2/4) \omega + 9zV^2 = 0 \: .
\end{equation}
The solutions for $V=0$ are easily obtained. For small $V\ne 0$ we find one
root near $\omega=0$:
\begin{equation}
   \omega_1(z) = \frac{36z}{U^2} \: V^2 + {\cal O}(V^4) \: ,
\end{equation}
and another one near $\omega=-U/2$:
\begin{eqnarray}
   \omega_2(z) &=& \frac{z}{2} - \frac{1}{2} \sqrt{z^2 + U^2} 
\nonumber \\ &-& 
   \: \left(
   \frac{18z}{U^2} + \frac{18}{U^2} \frac{z^2+U^2/2}{\sqrt{z^2+U^2}}
   \right) V^2 + {\cal O}(V^4) \: . \nonumber \\ 
\end{eqnarray}
Because of the step function in Eq.\ (\ref{eq:uclatt}), the third root near
$\omega=U/2$ is not needed here.
This yields:
\begin{eqnarray}
   && T \sum_\omega {\rm tr} \: \ln (-{\bm G}(i\omega)) / L = -R_\Sigma / L +
\nonumber \\ &&
   2 \int_{-\infty}^\infty dz \: \rho_0(z) \: 
   \left( \Theta(z) \frac{36z}{U^2} \: V^2 + \omega_2(z) \right)
   + {\cal O}(V^4) \: . 
   \nonumber \\
\label{eq:uclatt1}
\end{eqnarray}

Inserting the results, Eqs.\ (\ref{eq:ucie}), (\ref{eq:ucimp}), and (\ref{eq:uclatt1}),
into Eq.\ (\ref{eq:om}) and using the symmetry $\rho_0(z) = \rho_0(-z)$, we find:
\begin{eqnarray}
  \Omega / L &=& \mbox{const.} + V^2 \Bigg( 
  \frac{24}{U} + \frac{72}{U^2} \int_{-\infty}^0 dz \: \rho_0(z) \: z
  \nonumber \\ &-&
    \frac{36}{U^2} \int_{-\infty}^\infty dz \: \rho_0(z)
  \frac{z^2+U^2/2}{\sqrt{z^2+U^2}}  
  \Bigg) + {\cal O}(V^4) \: .  
  \nonumber \\
\end{eqnarray}
Now, the condition (\ref{eq:cond}) gives the critical $U$ for the Mott transition:
\begin{equation}
  U_{\rm c} = - 3 \int_{-\infty}^0 dz \: \rho_0(z) \: z
  + \frac{3}{2} \int_{-\infty}^\infty dz \: \rho_0(z)
  \frac{z^2+U_{\rm c}^2/2}{\sqrt{z^2+U_{\rm c}^2}} \: .
\label{eq:uc}  
\end{equation}
This implicit analytical equation for $U_{\rm c}$ is the final result.
For an arbitrary free DOS no further simplifications are
possible.

\section{Discussion}
\label{sec:results}

It should be stressed once more that Eq.\ (\ref{eq:uc}) results from 
an exact variational principle simply by the restriction that the trial 
self-energies be representable by the two-site reference system.
The two-site model generally yields a two-pole self-energy which is 
the minimal requirement for a three-peak structure of the single-particle 
excitation spectrum.
Therefore, it may also be stated that Eq.\ (\ref{eq:uc}) gives the optimal 
result for a two-pole self-energy. 

Eq.\ (\ref{eq:uc}) turns out to be more complicated as compared to
the result of the linearized DMFT: \cite{BP00}
\begin{equation}
  U^{\rm (L-DMFT)}_{\rm c} = 6 \; \sqrt{\int_{-\infty}^\infty dz \: \rho_0(z) \: z^2} \; .
\end{equation}
Interestingly, the first term on the r.h.s.\ of Eq.\ (\ref{eq:uc}) resembles
Brinkman and Rice (Gutzwiller) result \cite{BR70} for the critical interaction:
\begin{equation}
 U^{\rm (BR)}_{\rm c} = -16 \int_{-\infty}^0 dz \: \rho_0(z) \: z \; .
\end{equation}
A rough approximation of the second term in (\ref{eq:uc}) using
${(z^2+U_{\rm c}^2/2)}/{\sqrt{z^2+U_{\rm c}^2}} \to U_{\rm c}/2$
then yields
$U_{\rm c} \approx 3 U^{\rm (BR)}_{\rm c} / 4$.
This reduction of the critical interaction as compared to the BR
result is the dominating effect.

For a semi-elliptical free DOS with variance $\Delta = 1$
(band width $W=4$), one has $U^{\rm (L-DMFT)}_{\rm c}=6$ and
$U^{\rm (BR)}_{\rm c} = 6.7906$.
The numerical solution of Eq.\ (\ref{eq:uc}) is straightforward and 
yields $U_{\rm c} = 5.8450$.
As expected this is fully consistent with the numerical determination 
of $U_{\rm c}$ (Sec.\ \ref{sec:twosite}).
The result is close to the one predicted by the L-DMFT.
Compared to numerical estimates from full DMFT calculations
$U_{\rm c}^{\rm (DMFT)} = 5.84$ (NRG, Ref.\ \onlinecite{Bul99}) and
$U_{\rm c}^{\rm (DMFT)} = 5.88$ (PSCM, Ref.\ \onlinecite{MSK+95}), 
one can state that the prediction of the L-DMFT is improved.

\begin{figure}[t]
\centerline{\includegraphics[width=\columnwidth]{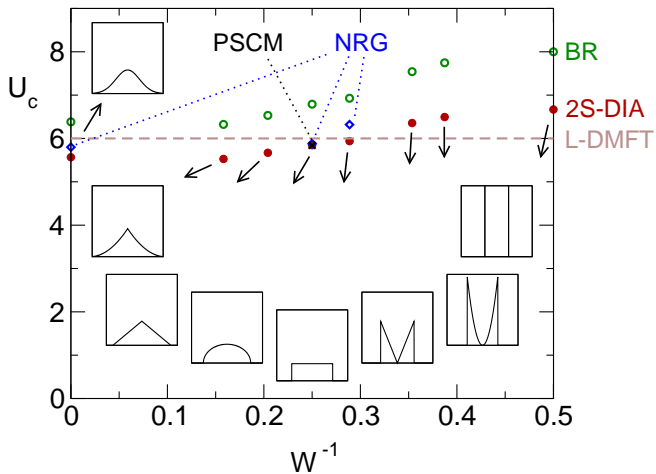}}
\caption{
Critical interaction $U_{\rm c}$ as a function of the inverse band width $W^{-1}$.
Calculations for different (normalized) free densities of states $\rho_0(z)$ 
as indicated (where $-3<z<3$ and $0<\rho_0(z)<1.5$ in the insets).
}
\label{fig:uc}
\end{figure}

\begin{table}[t]
\caption{
Critical interaction $U_{\rm c}$ as obtained from the SFA within the
$n_{\rm s}=2$ dynamical impurity approximation for different free
densities of states.
Each DOS is normalized, 
$\int_{-\infty}^\infty \rho_0(z) = 1$, symmetric,
$\rho_0(z) = \rho_0(-z)$, and has unit variance, 
$\Delta=\int_{-\infty}^\infty z^2 \rho_0(z)=1$.
$A(z)=\Theta(W/2+z)\Theta(W/2-z)$ is a cutoff function.
$W$ is the band width.
The critical interaction from the Gutzwiller (Brinkman-Rice) approach
\cite{BR70} and the available full DMFT results are shown for 
comparison. The linearized DMFT \cite{BP00} yields 
$U^{\rm (L-DMFT)}_{\rm c} = 6$ for any DOS with unit
variance.
\label{tab:uc}
}
\begin{ruledtabular}
\begin{tabular}{ccccc}
$\rho_0(z)$ & $W$ & $U_{\rm c}^{n_{\rm s}=2}$ & $U_{\rm c}^{\rm BR}$ & $U_{\rm c}^{\rm DMFT}$\\
\hline
Gaussian & $\infty$ & 5.5663 & 6.3831  & 5.80\footnotemark[1]
\\
$3A(z) (|z|-\sqrt{10})^2/20\sqrt{10}$  & $2\sqrt{10}$ &   $5.5284$ & 6.3246 &
\\
triangle & $2\sqrt{6}$ & 5.6696 & 6.5320 &
 \\
semi-ellipse & $4$ & $5.8450$ & 6.7906  & 5.84\footnotemark[2],5.88\footnotemark[1]
 \\
rectangle & $2\sqrt{3}$ & $5.9385$ & 6.9282  & 6.32\footnotemark[3]
 \\
$A(z) |z|/2$ & $2\sqrt{2}$ & $6.3554$ & 7.5425  &
 \\
$9\sqrt{3}A(z) z^2/10\sqrt{5}$ & $2\sqrt{5/3}$ & $6.4944$ & 7.7460  &
 \\
$\delta(z-1)/2+\delta(z+1)/2$ & $2$ & $6.6686$ & 8.0000  &
 \\
\end{tabular}
\end{ruledtabular}
\footnotetext[1]{Ref.~\onlinecite{Bul99}.}
\footnotetext[2]{Ref.~\onlinecite{MSK+95}.}
\footnotetext[3]{R. Bulla, private communication.}
\end{table}

Results for different free densities of states are displayed in 
Fig.\ \ref{fig:uc} and Tab.\ \ref{tab:uc}.
For a meaningful comparison, each DOS has unit variance 
$\Delta = 1$ where $\Delta^2 = \int_{-\infty}^\infty dz \rho_0(z) z^2$.
The band width $W$ varies.
The DOS with the smallest $W$ (but $\Delta = 1$) consists
of two $\delta$-peaks while $W = \infty$ for a Gaussian DOS.
Clearly, there is no true Mott transition in the former case 
(but also in the latter this is questionable).
However, the inclusion of these extreme cases is instructive when studying
the trend of $U_{\rm c}$ as a function of $W$.
Note that for $\Delta = 1$ the L-DMFT yields $U_{\rm c}=6$ irrespective of
the form of the DOS.

The critical interaction from the two-site DIA is always close the L-DMFT
result but considerably lower than the Gutzwiller value.
The two-site DIA confirms the central prediction of the L-DMFT that it is 
the variance of the DOS that is crucial for the critical interaction.
However, there is also a weak trend superimposed, namely a systematic 
increase of $U_{\rm c}$ with decreasing band width $W$ (with the exception 
of the Gaussian DOS). 
This is the same trend that is also present in the Gutzwiller results.
It would be interesting to see whether or not this trend is confirmed
by full DMFT calculations.
Comparing the full DMFT results for the Gaussian, for the rectangular and 
the semi-elliptic DOS shows the mentioned trend.
However, the comparison of the two-site DIA and of the L-DMFT
with the full DMFT for the available numerical data is not fully conclusive.

\section{Finite temperatures}
\label{sec:temp}

So far only the zero-temperature limit has been considered.
The Mott transition at finite temperatures, however, is particularly 
interesting as there is a comprehensive physical picture available 
with a comparatively complex phase diagram.
This phase diagram in the $U$-$T$ plane was first suggested by the 
iterative perturbation theory. \cite{GK92a,GKKR96} 
The nature of the transition and the topology of the phase diagram have 
been established entirely using analytic arguments \cite{Kot99} 
and has been worked out quantitatively using different numerical 
methods. \cite{RCK99,JO01,BCV01,Blu02}
The critical regime is accessible by the projective self-consistent
method. \cite{MSK+95}

As this phase diagram represents a valuable benchmark for any approximation,
it is interesting to see whether or not it can be rederived within the most 
simple two-site DIA.
The application of the theory for finite $T$ is straightforward but can
no longer be done analytically.
As for the derivation of the $T=0$ results in the non-critical regime 
(see Sec.\ \ref{sec:twosite}), calculations are performed along the
lines of Sec.\ \ref{sec:local}.

\begin{figure}[t]
\centerline{\includegraphics[width=0.75\columnwidth]{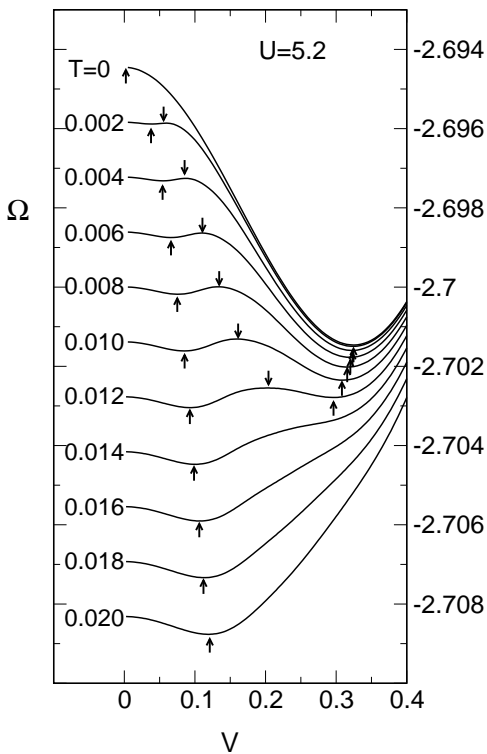}}
\caption{
Grand potential per site at $U=5.2$ for different $T$ 
as a function of $V$. 
The arrows indicate the stationary points.
}
\label{fig:omtemp}
\end{figure}

It turns out that for finite temperatures the Mott transition is 
predicted to be discontinuous.
This is demonstrated in Fig.\ \ref{fig:omtemp} which shows $\Omega(V)$ 
for different $T$ and fixed $U=5.2$ ($n_{\rm s}=2$).
At low but finite $T$ there are three stationary points (see arrows)
corresponding to three different phases of the system. 
The metallic phase has the largest value $V=V_{\rm met}$.
With increasing temperature $V_{\rm met}$ decreases and 
$\Omega(V_{\rm met}) \propto T^2$ for low $T$.
This gives a linear entropy $S(T) = - \partial \Omega (T) / \partial T$
and a linear specific heat $C_v = T \partial S(T) / \partial T
= \gamma T \propto z^{-1} T$ as it is characteristic for a Fermi liquid 
($\mu=U/2$ is fixed).
The insulating phase has the smallest value $V=V_{\rm ins}$, 
and the grand potential $\Omega(V_{\rm ins}) \propto T$ for low $T$.
For $T\to 0$ the entropy approaches $S \to L \ln 2$ reflecting a $2^L$-fold 
ground-state degeneracy of the insulator which is known to be an artifact of 
mean-field theory. \cite{GKKR96} 
The specific heat $C_v$ vanishes exponentially for $T \to 0$.
For $U=5.2$ and low $T$, however, the insulating phase is metastable
as compared to the metallic phase since
$\Omega(V_{\rm ins}) > \Omega(V_{\rm met})$.
Due to the different behavior at low $T$, there must be a temperature
$T_c(U)$ where $\Omega(V_{\rm ins}) = \Omega(V_{\rm met})$.
In fact, for $T = 0.012 > T_c(U=5.2)$ the insulator is stable as compared 
to the metal.
At $T_c(U)$ or, conversely, at a critical interaction $U_c(T)$, the 
$n_{\rm s}=2$-DIA thus predicts a first-order transition with
a discontinuous jump in the entropy.
As can be seen in Fig.\ \ref{fig:omtemp}, the metallic phase ceases to 
exist for still higher temperatures, and only the insulating phase is
left. This is due to a coalescence of the metallic with a third phase at another
critical temperature $T_{c2}(U)$ (or, conversely, at a critical interaction
$U_{c2}(T))$. This third phase turns out to be less stable as compared
to the metal and to the insulator in the entire parameter regime.
Similarly, one can define a critical temperature $T_{c1}(U)$ (at smaller
$U$) and thus a critical interaction $U_{c1}(T)$ where the insulating 
phase coalesces with the third phase.

\begin{figure}[t]
\centerline{\includegraphics[width=0.9\columnwidth]{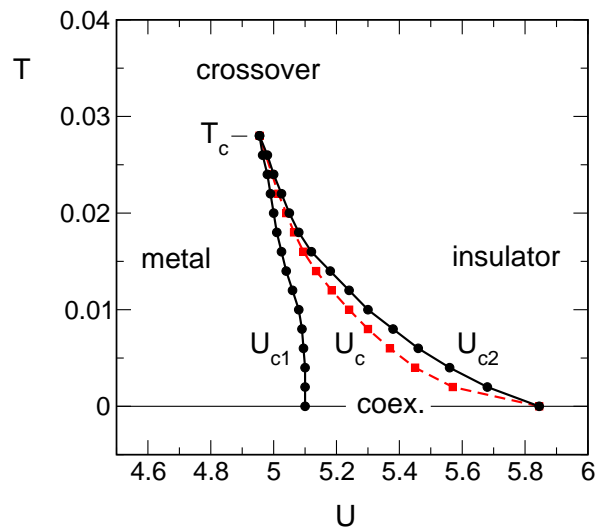}}
\caption{
Phase diagram for the Mott transition in the paramagnetic phase of the
Hubbard model at half-filling. Calculations for a semi-elliptical density
of states with band width $W=4$. 
Reference system $H'$: two-site model.
$U_{c2}$: critical interaction strength up to which there is a (metallic) 
solution smoothly connected with the metal at $U=0$.
$U_{c1}$: critical interaction strength down to which there is an (insulating)
solution smoothly connected with the Mott insulator for $U\to \infty$.
$U_c$: first-order transition line in the coexistence region, terminating at
the second-order critical point at $T_c$.
}
\label{fig:pd}
\end{figure}

Calculations for different $U$ and $T$ have been performed to obtain
the entire phase diagram within the two-site approach. 
The result is shown in Fig.\ \ref{fig:pd}.
For $U \le U_{c2}(T)$ there is a metallic phase which is smoothly 
connected to the $U=0$ limit.
On the other hand, for 
$U \ge U_{c1}(T)$ there is an insulating phase which is smoothly connected 
with the Mott insulator for $U\to \infty$.
Metallic and insulating phase are coexisting for $U_{c1}(T) \le U \le U_{c2}(T)$.
At zero temperature, the metal is stable as compared to the insulator in the
entire coexistence region, and the transition is continuous at $U_c = U_{c2}$.
For finite temperatures the transition is discontinuous at a critical interaction
$U_c(T)$ with $U_{c1}(T) \le U_c(T) \le U_{c2}(T)$. 
With increasing temperatures the coexistence region $U_{c1}(T) \le U \le U_{c2}(T)$
shrinks and disappears above a critical temperature $T_c$ which is defined via
$U_{c1}(T_c) = U_{c2}(T_c)$.
Above $T_c$ the metallic phase is smoothly connected with the insulating phase.

Qualitatively, this is exactly the same result that is obtained within 
the full DMFT. \cite{GKKR96}
It is very remarkable that the rather complex topology of the phase diagram
can be reproduced with a comparatively simple two-site model as a reference 
system.
This shows that it is essential to perform the mapping onto the reference
system in a way which is thermodynamically consistent and which is controlled 
by a physical variational principle.
Note that the L-DMFT, or its extension away from the critical point at $T=0$, 
\cite{Pot01} fails to reproduce the discontinuous transition for $T>0$ and 
the critical temperature $T_c$ due to the ad-hoc character of the 
approximation.

Quantitatively, one should expect some deviations from the results of the 
full DMFT due to the simplicity of the two-site model. 
Comparing with the NRG result \cite{BCV01} for $U_{c1}(T=0) \approx 4.8 - 5.0$, 
the two-site approximation overestimates the critical interaction by a few
per cent.
The determination of the critical temperature is difficult in any numerical 
approach. 
$T_c \approx 0.05 - 0.08$ is estimated from the QMC and NRG results of Refs.\ 
\onlinecite{RCK99,BCV01}.
Thus, the two-site approximation underestimates $T_c$ by more than a 
factor 2.
It is worth mentioning that the numerical effort to obtain the entire phase 
diagram in the $U-T$ plane is of the order of a few minutes on a standard
workstation which is negligible as compared to a DMFT-QMC calculation.

\section{Conclusions}
\label{sec:con}

Our present understanding of the Mott-Hubbard transition is mainly based
on the exact solution of the one-dimensional case on the one hand and on
the $D=\infty$ mean-field picture provided by the DMFT on the other hand.
For the physically more relevant two- or three-dimensional Hubbard model,
however, neither the analytical concepts developed for $D=1$ nor the 
mean-field theory can be expected to give an essentially correct and
comprehensive description.
Direct numerical approaches, such as QMC, are able to give essentially 
exact results for a $D=2,3$ dimensional lattice of finite spatial extent. 
However, the relevance of the results for the thermodynamic limit and, 
in many cases, for the low-temperature, low-energy regime remains an 
open question.

In this situation, a combination of a direct numerical approach for systems
of finite size with the mean-field concept appears to be advantageous.
This is more or less the direction that is followed up by the different 
cluster extensions of the DMFT. \cite{SI95,HTZ+98,LK00,KSPB01}
Via a generalized mean-field concept, the original lattice problem given 
by the Hamiltonian $H$ is mapped onto a cluster problem described by a 
Hamiltonian $H'$.
In fact, due to the presence of strong short-range antiferromagnetic 
correlations a considerable revision of the mean-field picture is probably 
necessary. \cite{MJPK00}
However, the cluster extensions of the DMFT suffer from the fact that
the mean-field formulation requires that the sites of the finite cluster
are coupled to uncorrelated baths with an infinite number of degrees
of freedom each.
This circumstance complicates the practical solution of the $H'$ problem
(which must be solved repeatedly during the self-consistency cycle) so much 
that additional approximations are required and/or stochastic numerical 
methods. 

The self-energy-functional approach offers an interesting alternative as 
the reference system $H'$, the original model $H$ is mapped onto, is by
no means completely predetermined.
The SFA gives a very general prescription how this mapping can be performed
while keeping the thermodynamical consistency of the approach as an explicit
expression for the grand potential is provided.
In this way the DMFT and the cellular DMFT are recovered as certain 
limits for special choices of $H'$, namely for a decoupled system of 
clusters of size $N_{\rm c}=1$ or $N_{\rm c}>1$ including couplings to 
$n_{\rm b}=\infty$ bath degrees of freedom.
There is, however, the additional possibility to construct approximations
with $n_{\rm b} < \infty$ which are consistent in themselves in the same 
way as are the DMFT and the C-DMFT.

Now, the question is whether or not an approximation with a finite number
of bath sites $n_{\rm b} < \infty$ can be tolerated.
Note that even in the (C-)DMFT it becomes necessary to reduce the problem 
posed by $H'$ with $n_{\rm b}=\infty$ to a numerically tractable one with 
a {\em finite} number of degrees of freedom.
While this additional approximation is controlled within QMC or ED 
approaches, for example, it nevertheless violates thermodynamical 
consistency.
Within the SFA, on the contrary, the approximation is derived from a 
thermodynamical potential for any $n_{\rm b}$, infinite or finite or 
even so small as $n_{\rm b} = 1$.
Depending on the quantity and accuracy one is interested in, there can 
be a rapid convergence with respect to $n_{\rm b}$ 
(cf.\ Ref.\ \onlinecite{Pot03a}) so that a small number might be sufficient.
For cluster approximations, the best choice is by no means clear as it must 
be balanced with the choice of the cluster size $N_{\rm c}$. 
This strongly depends on the lattice dimension.
In fact, it has been shown \cite{PAD03} that for the one-dimensional 
Hubbard model a larger $N_{\rm c}$ is to be preferred as compared to 
a larger $n_{\rm b}$.

In this context, it is interesting to see where one are led to with the 
most simple reference system conceivable. 
This is a model $H'$ characterized by $N_{\rm c} = 1$ and $n_{\rm b} = 1$ 
which yields the so-called two-site dynamical impurity approximation (DIA).
The answer is given with the present paper:
Even in this approximation the Mott transition shows up.
At zero temperature the transition turns out to be continuous at a 
finite critical interaction $U_{\rm c}$ where there is a coalescence of 
the metallic with a coexisting insulating phase.
For finite temperatures, on the other hand, the transition is discontinuous.
The first-order line $U_c(T)$ terminates at a second-order critical point
($U_c(T_c), T_c$), and for $T>T_c$ there is a smooth crossover only.
This is qualitatively the same picture that has been found beforehand in 
the full DMFT.
Furthermore, the two-site DIA yields a $U_{\rm c}$ at $T=0$ that is surprisingly 
close to the result of the full DMFT.
In this respect the two-site DIA is of similar quality as the linearized 
DMFT, \cite{BP00} another approach that is based on a mapping onto the 
two-site SIAM.
Whether or not the $U_{\rm c}$-results of the two-site DIA improve on 
those of the L-DMFT is difficult to decide in view of the existing (full) 
DMFT data.
More important, however, the conceptual improvement gained is substantial:
While the mapping procedure on the two-site model, though physically 
motivated, is done in an ad-hoc way in the L-DMFT, the two-site DIA is 
derivable from a thermodynamical potential and can be characterized as 
an optimal two-site approach in fact.

The fact that a reasonable mean-field description of the Mott transition 
is possible even for the simple two-site DIA, motivates further SFA studies
of the transition with improved approximations in the future (e.g.\ using 
finite clusters, $N_{\rm c} > 1$, and small $n_{\rm b}$).
In case of larger and more complex reference systems, however, it becomes 
more and more important to have a numerically efficient and accurate 
method for the evaluation of the self-energy functional at hand.
The detailed analysis of the different contributions to the functional
has been another major intention of the present study.
Using the causality properties of the Green function, it has been shown 
that the important ${\rm Tr} \ln (- {\bm G} )$ term can be written as
the grand potential of a system of non-interacting quasi-particles with
unit weight (apart from a correction term that cancels out in the 
functional eventually).
While this result is also interesting by itself, it is very well suited 
for an efficient numerical implementation.
As the energies of the ficticious quasi-particles are given by the poles
of the Green function, they can be found in a comparatively simple way
by exploiting general causality properties.
In particular, there is no need for a small but finite Lorentzian 
broadening, $\omega \to \omega + i \delta$, which must be introduced in
a more direct way to evaluate the functional. \cite{PAD03,DAH+03}
This will become important when studying the critical regime of the Mott 
transition, where an accurate computation of small energy differences is 
vital, using an approximation with several variational parameters.
Studies in this direction are intended for the future.

\begin{acknowledgments}
It is a pleasure to acknowledge valuable discussions with 
B. Michaelis, E. Arrigoni and W. Hanke.
This work is supported by the Deutsche Forschungsgemeinschaft 
(SFB~290 
``Metallische d\"unne Filme: Struktur, Magnetismus und elektronische Eigenschaften''
and SFB~410
``II-VI-Halbleiter: Wachstumsmechanismen, niederdimensionale Strukturen und Grenzfl\"achen'').
\end{acknowledgments}

\appendix

\section{}
\label{sec:causal}

Here is will be shown that the Green's function ${\bm G}$ is causal:
For any ${\bm t}'$ the self-energy ${\bm \Sigma}={\bm \Sigma}({\bm t}')$ 
is causal since it is defined to be the exact self-energy of the reference 
system $H' = H_0({\bm t}') + H_1 ({\bm U})$. 
Likewise, the Green's function ${\bm G}_0$ is causal.
It has to be shown that ${\bm G} \equiv ({\bm G}_0^{-1} - {\bm \Sigma})^{-1}$
is causal if ${\bm \Sigma}$ and the free Green's function ${\bm G}_0$ are known 
to be causal.

Causality of ${\bm G}$ means (i) that $G_{\alpha\beta}(\omega)$ is analytic
in the entire complex $\omega$ plane except for the real axis and (ii) that
${\bm G}_{\rm ret}(\omega) = 
{\bm G}(\omega + i0^+) = {\bm G}_{\rm R} - i {\bm G}_{\rm I}$ for real $\omega$ with 
${\bm G}_{\rm R}$, ${\bm G}_{\rm I}$ Hermitian and ${\bm G}_{\rm I}$ 
positive definite.
(i) is easily verified. To show (ii) we need the following

{\em Lemma:} For Hermitian matrices ${\bm A}$, ${\bm B}$ with
${\bm B}$ positive definite, one has
\begin{equation}
   \frac{1}{{\bm A} \pm i {\bm B}} = {\bm X} \mp i {\bm Y}
\end{equation}
with ${\bm X}$, ${\bm Y}$ Hermitian and ${\bm Y}$ positive definite.
The {\em proof} of the Lemma is straightforward:
\begin{equation}
   \frac{1}{{\bm A} \pm i {\bm B}} = 
   {\bm B}^{-\frac{1}{2}} \,
   \frac{1}{{\bm B}^{-\frac{1}{2}} {\bm A} {\bm B}^{-\frac{1}{2}} \pm i} \,
   {\bm B}^{-\frac{1}{2}} 
   = {\bm D} \,\frac{1}{{\bm C} \pm i} \,{\bm D}
\end{equation}
with ${\bm C}$, ${\bm D}$ Hermitian and 
${\bm D}={\bm B}^{-\frac{1}{2}}$ positive definite.
Let ${\bm C} = {\bm U} {\bm c} {\bm U}^\dagger$ with ${\bm U}$ unitary
and ${\bm c}$ diagonal. Then
\begin{equation}
   \frac{1}{{\bm A} \pm i {\bm B}} 
   = {\bm D} {\bm U} \,\frac{1}{{\bm c} \pm i} \, {\bm U}^\dagger {\bm D}
   = {\bm D} {\bm U} \,\frac{{\bm c} \mp i}{{\bm c}^2 + 1} \, {\bm U}^\dagger {\bm D}
   = {\bm X} \mp i {\bm Y}
\end{equation}
with ${\bm X}$ Hermitian and 
\begin{equation}
  {\bm Y} = 
  {\bm D} {\bm U} \,\frac{1}{{\bm c}^2 + 1} \, {\bm U}^\dagger {\bm D}
\end{equation}
Hermitian and positive definite.

Consider a fixed (real) frequency $\omega$.
Since ${\bm G}_{0}$ is causal, 
${\bm G}_{0,{\rm ret}} = {\bm G}_{0, \rm R} - i {\bm G}_{0, \rm I}$ with 
${\bm G}_{0, \rm R}$, ${\bm G}_{0, \rm I}$ Hermitian and 
${\bm G}_{0, \rm I}$ positive definite.
Using the lemma, 
${\bm G}_{0,{\rm ret}}^{-1} = {\bm P}_{\rm R} + i {\bm P}_{\rm I}$
with ${\bm P}_{\rm R}$, ${\bm P}_{\rm I}$ Hermitian and ${\bm P}_{\rm I}$ 
positive definite.
Since ${\bm \Sigma}$ is causal, 
${\bm \Sigma}_{\rm ret} = {\bm \Sigma}_{\rm R} - i {\bm \Sigma}_{\rm I}$ 
with ${\bm \Sigma}_{\rm R}$, ${\bm \Sigma}_{\rm I}$ Hermitian and 
${\bm \Sigma}_{\rm I}$ positive definite. Therefore,
\begin{equation}
  {\bm G}_{\rm ret} 
  = \frac{1}{{\bm P}_{\rm R} + i {\bm P}_{\rm I} 
  - {\bm \Sigma}_{\rm R} + i {\bm \Sigma}_{\rm I}}
  = \frac{1}{{\bm Q}_{\rm R} + i {\bm Q}_{\rm I}}
\end{equation}
with ${\bm Q}_{\rm R}$ Hermitian and ${\bm Q}_{\rm I}$ Hermitian 
and positive definite.
Using the lemma once more, shows ${\bm G}$ to be causal.

\section{}
\label{sec:fom}

Consider the function
\begin{equation}
  f(\omega) = \sum_m \frac{R_m}{\omega - \omega_m}
\label{eq:pos1}
\end{equation}
with real isolated first-order poles at $\omega = \omega_m$ and
positive residues $R_m > 0$.
We have:
\begin{equation}
  -\frac{1}{\pi} \mbox{Im} f(\omega+i0^+) > 0
\label{eq:pos2}
\end{equation}
In Sec.\ \ref{sec:eval} $f(\omega) = g_k(\omega) = 1/(\omega + \mu - \eta_k(\omega))$.

In a polar representation $-f(\omega+i0^+) = r(\omega) e^{i\phi(\omega)}$ with
$-\pi < \phi(\omega) \le \pi$.
On the principal branch of the logarithm and for real $\omega$ one has
$\mbox{Im} \ln (-f(\omega+i0^+)) = \phi(\omega)$.
For any $\omega \ne \omega_m$ we have $\mbox{Im} (- f(\omega+i0^+)) = 0^+$ from
Eqs.\ (\ref{eq:pos1}) and (\ref{eq:pos2}) and
thus $\phi(\omega) = \pi$ for $-f(\omega) < 0$, and 
$\phi(\omega) = 0$ for $-f(\omega) > 0$. 
Consequently,
\begin{equation}
  {\rm Im} \ln ( - f(\omega+i0^+)) = \pi \, \Theta(f(\omega)) 
  = \pi \, \Theta(1/f(\omega)) \: , 
\end{equation}
where $\Theta$ is the step function.
For $\omega = \omega_m$, the imaginary part of $-f(\omega+i0^+)$ diverges. 
However, $-\pi < \phi(\omega) \le \pi$ 
(in fact $\phi(\omega_m+0^+) = -\pi/2$ and $\phi(\omega_m-0^+) = \pi/2$).
Hence, ${\rm Im} \, \ln ( - f(\omega+i0^+))$ remains finite at $\omega=\omega_m$
and can be ignored in an integration over real $\omega$ as the poles of $f(\omega)$ 
are isolated.

\section{}
\label{sec:theta}

Consider a function $f=f(\omega)$ which is analytical except for isolated 
first-order poles on the real axis and which is real for real $\omega$.
$\Theta(f(\omega))$ is constant almost everywhere, and thus the derivative
$d \Theta(f(\omega)) / d \omega$ vanishes almost everywhere. 
A non-zero derivative may either occur if $f(\omega)=0$, i.~e.\ at the zeros 
$\omega_m$. This gives 
a contribution:
\begin{equation}
  \delta(f(\omega)) f'(\omega) = 
  \sum_m \frac{ f'(\omega_m) }{ \left| f'(\omega_m) \right| } \delta(\omega-\omega_m)
\end{equation}
where $f'(\omega)\equiv df(\omega)/d\omega$. 
A second contribution arises from
the first-order poles of $f(\omega)$ at $\zeta_n$. The poles are
the zeros of the function $1/f(\omega)$. Note that 
$\Theta(f(\omega))=\Theta(1/f(\omega))$ since the sign of the argument is
unchanged. Thus the contribution due to the poles is:
\begin{equation}
  \delta(1/f(\omega)) (1/f)'(\omega) = 
  \sum_n \frac{(1/f)'(\zeta_n)}{\left| (1/f)'(\zeta_n) \right|} 
  \delta(\omega-\zeta_n)
\end{equation}
We have:
\begin{eqnarray}
   \frac{d \Theta(f(\omega)) }{d \omega} &=&
   \delta(f(\omega)) f'(\omega) + 
   \delta(1/f(\omega)) (1/f)'(\omega)
\nonumber \\ &=&
   \sum_m^{\{\rm zeros\}} \epsilon_m \delta(\omega-\omega_m)
   + \!\!
   \sum_n^{\{\rm poles\}} \epsilon_n \delta(\omega-\zeta_n)
\nonumber \\
\end{eqnarray}
with $\epsilon_m=\pm 1$ and $\epsilon_n=\pm 1$ depending on the sign of the 
slope of $f$ at the zeros $\omega_m$ and the sign of the residue of
$f$ at the poles $\zeta_n$, respectively.

Consider the function $f(\omega)=\omega + \mu - \eta_k(\omega)$.
The zeros of $f$ are the poles of the diagonalized Green's function
$1/(\omega + \mu - \eta_k(\omega))$ which has positive residues.
This implies a positive $f'$ at $\omega_m$ and $\epsilon_m=+1$. 
The poles of $f$ are the poles of the self-energy which has positive 
residues. Thus the residues of $f$ at the poles are negative and
$(1/f)'$ is negative at $\zeta_n$ and $\epsilon_n=-1$. 
Hence:
\begin{eqnarray}
   \frac{d}{d \omega} \Theta(\omega + \mu - \eta_k(\omega)) =
   \sum_m \delta(\omega-\omega_m)
   -
   \sum_n \delta(\omega-\zeta_n) \: .
\nonumber \\
\end{eqnarray}

\section{}
\label{sec:hur}

The reference system $H'$ is a set of decoupled single-impurity Anderson 
models with one-particle energy of the impurity site $\epsilon_1$, conduction-band
energies $\epsilon_k$ with $k=2,...,n_{\rm s}$ and hybridization strengths
$V_k$.
We first calculate the eigenvalues of the hopping matrix ${\bm t}'$.
The matrix is block diagonal with respect to the site index $i$.
Each block is labeled by the ``orbital'' index $k=1,...,n_{\rm s}$.
There are non-zero elements of the matrix for $k=1$, $k'=1$, and
$k=k'$ only: 
\begin{equation}
  t'_{kk'} = (1-\delta_{k1})\delta_{k'1} \; V_{k} 
  + \delta_{k1}(1-\delta_{k'1}) \;  V_{k'} + \delta_{kk'} \; \epsilon_{k'}
\end{equation}
Furthermore, $\Sigma_{kk'}(\omega) = \delta_{k1} \delta_{k'1} \Sigma(\omega)$.
Using 
\begin{eqnarray}
\det \left(
\begin{array}{ccccc}
  d_1 & a^\ast_2 & a^\ast_3 & a^\ast_4 & ... \\
  a_2 & d_2 & 0 & 0 & ... \\
  a_3 & 0 & d_3 & 0 & ... \\
  a_4 & 0 & 0 & d_4 & ... \\ 
  ... &...&...& ... & ...
  \end{array}
\right)
=
\det \left(
\begin{array}{ccccc}
  \widetilde{d}_1 & 0 & 0 & 0 & ... \\
  0 & d_2 & 0 & 0 & ... \\
  0 & 0 & d_3 & 0 & ... \\
  0 & 0 & 0 & d_4 & ... \\ 
  ... &...&...& ... & ...
  \end{array}
\right)
\nonumber \\
\end{eqnarray}
where $\widetilde{d}_1 = d_1 - \sum_{k=2}^{n_{\rm s}} | a_k |^2 / d_k$
and the general relation ${\rm tr} \ln {\bm A} = \ln \det {\bm A}$, we find:
\begin{eqnarray}
  \frac{1}{2} {\rm tr}' \ln (-{\bm G}'(i\omega)) 
  &=& \ln \det \frac{-1}{i\omega + \mu - {\bm t}' - {\bm \Sigma}(i\omega)}
\nonumber \\
  &=& \ln \det 
  \left(
  \begin{array}{cccc}
  - G'_1 &0&0&... \\
  0&- G'_2 &0&... \\
  0&0&- G'_3 &... \\
  ...&...&...& ... 
  \end{array}
  \right)
\nonumber \\
  &=& \ln (- G_1(i\omega)) + \sum_{k=2}^{n_{\rm s}} \ln (- G_k(i\omega))
\nonumber \\
\end{eqnarray}
with $G_1(i\omega)$ and $G_k(i\omega)$ for $k=2,...,n_{\rm s}$ as defined
by Eqs.\ (\ref{eq:g1def}) and (\ref{eq:grdef}).
The factor $1/2$ accounts for the two spin directions.


\begin{thebibliography}{51}
\expandafter\ifx\csname natexlab\endcsname\relax\def\natexlab#1{#1}\fi
\expandafter\ifx\csname bibnamefont\endcsname\relax
  \def\bibnamefont#1{#1}\fi
\expandafter\ifx\csname bibfnamefont\endcsname\relax
  \def\bibfnamefont#1{#1}\fi
\expandafter\ifx\csname citenamefont\endcsname\relax
  \def\citenamefont#1{#1}\fi
\expandafter\ifx\csname url\endcsname\relax
  \def\url#1{\texttt{#1}}\fi
\expandafter\ifx\csname urlprefix\endcsname\relax\def\urlprefix{URL }\fi
\providecommand{\bibinfo}[2]{#2}
\providecommand{\eprint}[2][]{\url{#2}}

\bibitem[{\citenamefont{Imada et~al.}(1998)\citenamefont{Imada, Fujimori, and
  Tokura}}]{IFT98}
\bibinfo{author}{\bibfnamefont{M.}~\bibnamefont{Imada}},
  \bibinfo{author}{\bibfnamefont{A.}~\bibnamefont{Fujimori}}, \bibnamefont{and}
  \bibinfo{author}{\bibfnamefont{Y.}~\bibnamefont{Tokura}},
  \bibinfo{journal}{Rev. Mod. Phys.} \textbf{\bibinfo{volume}{68}},
  \bibinfo{pages}{13} (\bibinfo{year}{1998}).

\bibitem[{\citenamefont{Gebhard}(1997)}]{Geb97}
\bibinfo{author}{\bibfnamefont{F.}~\bibnamefont{Gebhard}},
  \emph{\bibinfo{title}{The Mott Metal-Insulator Transition}}
  (\bibinfo{publisher}{Springer}, \bibinfo{address}{Berlin},
  \bibinfo{year}{1997}).

\bibitem[{\citenamefont{Mott}(1990)}]{Mot90}
\bibinfo{author}{\bibfnamefont{N.~F.} \bibnamefont{Mott}},
  \emph{\bibinfo{title}{Metal-Insulator Transitions}}
  (\bibinfo{publisher}{Taylor and Francis}, \bibinfo{address}{London},
  \bibinfo{year}{1990}), \bibinfo{edition}{2nd} ed.

\bibitem[{\citenamefont{Hubbard}(1963)}]{Hub63}
\bibinfo{author}{\bibfnamefont{J.}~\bibnamefont{Hubbard}},
  \bibinfo{journal}{Proc. R. Soc. London A} \textbf{\bibinfo{volume}{276}},
  \bibinfo{pages}{238} (\bibinfo{year}{1963}).

\bibitem[{\citenamefont{Gutzwiller}(1963)}]{Gut63}
\bibinfo{author}{\bibfnamefont{M.~C.} \bibnamefont{Gutzwiller}},
  \bibinfo{journal}{Phys. Rev. Lett.} \textbf{\bibinfo{volume}{10}},
  \bibinfo{pages}{159} (\bibinfo{year}{1963}).

\bibitem[{\citenamefont{Kanamori}(1963)}]{Kan63}
\bibinfo{author}{\bibfnamefont{J.}~\bibnamefont{Kanamori}},
  \bibinfo{journal}{Prog. Theor. Phys. (Kyoto)} \textbf{\bibinfo{volume}{30}},
  \bibinfo{pages}{275} (\bibinfo{year}{1963}).

\bibitem[{\citenamefont{Luttinger and Ward}(1960)}]{LW60}
\bibinfo{author}{\bibfnamefont{J.~M.} \bibnamefont{Luttinger}}
  \bibnamefont{and} \bibinfo{author}{\bibfnamefont{J.~C.} \bibnamefont{Ward}},
  \bibinfo{journal}{Phys. Rev.} \textbf{\bibinfo{volume}{118}},
  \bibinfo{pages}{1417} (\bibinfo{year}{1960}).

\bibitem[{\citenamefont{Dagotto}(1994)}]{Dag94}
\bibinfo{author}{\bibfnamefont{E.}~\bibnamefont{Dagotto}},
  \bibinfo{journal}{Rev. Mod. Phys.} \textbf{\bibinfo{volume}{66}},
  \bibinfo{pages}{763} (\bibinfo{year}{1994}).

\bibitem[{\citenamefont{Georges and Kotliar}(1992)}]{GK92a}
\bibinfo{author}{\bibfnamefont{A.}~\bibnamefont{Georges}} \bibnamefont{and}
  \bibinfo{author}{\bibfnamefont{G.}~\bibnamefont{Kotliar}},
  \bibinfo{journal}{Phys. Rev. B} \textbf{\bibinfo{volume}{45}},
  \bibinfo{pages}{6479} (\bibinfo{year}{1992}).

\bibitem[{\citenamefont{Jarrell}(1992)}]{Jar92}
\bibinfo{author}{\bibfnamefont{M.}~\bibnamefont{Jarrell}},
  \bibinfo{journal}{Phys. Rev. Lett.} \textbf{\bibinfo{volume}{69}},
  \bibinfo{pages}{168} (\bibinfo{year}{1992}).

\bibitem[{\citenamefont{Georges et~al.}(1996))\citenamefont{Georges, Kotliar,
  Krauth, and Rozenberg}}]{GKKR96}
\bibinfo{author}{\bibfnamefont{A.}~\bibnamefont{Georges}},
  \bibinfo{author}{\bibfnamefont{G.}~\bibnamefont{Kotliar}},
  \bibinfo{author}{\bibfnamefont{W.}~\bibnamefont{Krauth}}, \bibnamefont{and}
  \bibinfo{author}{\bibfnamefont{M.~J.} \bibnamefont{Rozenberg}},
  \bibinfo{journal}{Rev. Mod. Phys.} \textbf{\bibinfo{volume}{68}},
  \bibinfo{pages}{13} (\bibinfo{year}{1996}).

\bibitem[{\citenamefont{Metzner and Vollhardt}(1989)}]{MV89}
\bibinfo{author}{\bibfnamefont{W.}~\bibnamefont{Metzner}} \bibnamefont{and}
  \bibinfo{author}{\bibfnamefont{D.}~\bibnamefont{Vollhardt}},
  \bibinfo{journal}{Phys. Rev. Lett.} \textbf{\bibinfo{volume}{62}},
  \bibinfo{pages}{324} (\bibinfo{year}{1989}).

\bibitem[{\citenamefont{M\"uller-Hartmann}(1989)}]{MH89a}
\bibinfo{author}{\bibfnamefont{E.}~\bibnamefont{M\"uller-Hartmann}},
  \bibinfo{journal}{Int. J. Mod. Phys. B} \textbf{\bibinfo{volume}{3}},
  \bibinfo{pages}{2169} (\bibinfo{year}{1989}).

\bibitem[{\citenamefont{Brandt and Mielsch}(1989)}]{BM89}
\bibinfo{author}{\bibfnamefont{U.}~\bibnamefont{Brandt}} \bibnamefont{and}
  \bibinfo{author}{\bibfnamefont{C.}~\bibnamefont{Mielsch}},
  \bibinfo{journal}{Z. Phys. B} \textbf{\bibinfo{volume}{75}},
  \bibinfo{pages}{365} (\bibinfo{year}{1989}).

\bibitem[{\citenamefont{Brandt and Mielsch}(1990)}]{BM90}
\bibinfo{author}{\bibfnamefont{U.}~\bibnamefont{Brandt}} \bibnamefont{and}
  \bibinfo{author}{\bibfnamefont{C.}~\bibnamefont{Mielsch}},
  \bibinfo{journal}{Z. Phys. B} \textbf{\bibinfo{volume}{79}},
  \bibinfo{pages}{295} (\bibinfo{year}{1990}).

\bibitem[{\citenamefont{Georges and Krauth}(1993)}]{GK93}
\bibinfo{author}{\bibfnamefont{A.}~\bibnamefont{Georges}} \bibnamefont{and}
  \bibinfo{author}{\bibfnamefont{W.}~\bibnamefont{Krauth}},
  \bibinfo{journal}{Phys. Rev. B} \textbf{\bibinfo{volume}{48}},
  \bibinfo{pages}{7167} (\bibinfo{year}{1993}).

\bibitem[{\citenamefont{Rozenberg
  et~al.}(1994){\natexlab{a}})\citenamefont{Rozenberg, Kotliar, and
  Zhang}}]{RKZ94}
\bibinfo{author}{\bibfnamefont{M.~J.} \bibnamefont{Rozenberg}},
  \bibinfo{author}{\bibfnamefont{G.}~\bibnamefont{Kotliar}}, \bibnamefont{and}
  \bibinfo{author}{\bibfnamefont{X.~Y.} \bibnamefont{Zhang}},
  \bibinfo{journal}{Phys. Rev. B} \textbf{\bibinfo{volume}{49}},
  \bibinfo{pages}{10181} (\bibinfo{year}{1994}{\natexlab{a}}).

\bibitem[{\citenamefont{Caffarel and Krauth}(1994)}]{CK94}
\bibinfo{author}{\bibfnamefont{M.}~\bibnamefont{Caffarel}} \bibnamefont{and}
  \bibinfo{author}{\bibfnamefont{W.}~\bibnamefont{Krauth}},
  \bibinfo{journal}{Phys. Rev. Lett.} \textbf{\bibinfo{volume}{72}},
  \bibinfo{pages}{1545} (\bibinfo{year}{1994}).

\bibitem[{\citenamefont{Rozenberg
  et~al.}(1994){\natexlab{b}})\citenamefont{Rozenberg, Moeller, and
  Kotliar}}]{RMK94}
\bibinfo{author}{\bibfnamefont{M.}~\bibnamefont{Rozenberg}},
  \bibinfo{author}{\bibfnamefont{G.}~\bibnamefont{Moeller}}, \bibnamefont{and}
  \bibinfo{author}{\bibfnamefont{G.}~\bibnamefont{Kotliar}},
  \bibinfo{journal}{Mod. Phys. Lett. B} \textbf{\bibinfo{volume}{8}},
  \bibinfo{pages}{535} (\bibinfo{year}{1994}{\natexlab{b}}).

\bibitem[{\citenamefont{Si et~al.}(1994)\citenamefont{Si, Rozenberg, Kotliar,
  and Ruckenstein}}]{SRKR94}
\bibinfo{author}{\bibfnamefont{Q.}~\bibnamefont{Si}},
  \bibinfo{author}{\bibfnamefont{M.~J.} \bibnamefont{Rozenberg}},
  \bibinfo{author}{\bibfnamefont{G.}~\bibnamefont{Kotliar}}, \bibnamefont{and}
  \bibinfo{author}{\bibfnamefont{A.~E.} \bibnamefont{Ruckenstein}},
  \bibinfo{journal}{Phys. Rev. Lett.} \textbf{\bibinfo{volume}{72}},
  \bibinfo{pages}{2761} (\bibinfo{year}{1994}).

\bibitem[{\citenamefont{Eastwood et~al.}(2003)\citenamefont{Eastwood, Gebhard,
  Kalinowski, Nishimoto, and Noack}}]{EGK+03}
\bibinfo{author}{\bibfnamefont{M.~P.} \bibnamefont{Eastwood}},
  \bibinfo{author}{\bibfnamefont{F.}~\bibnamefont{Gebhard}},
  \bibinfo{author}{\bibfnamefont{E.}~\bibnamefont{Kalinowski}},
  \bibinfo{author}{\bibfnamefont{S.}~\bibnamefont{Nishimoto}},
  \bibnamefont{and} \bibinfo{author}{\bibfnamefont{R.~M.} \bibnamefont{Noack}},
  \bibinfo{journal}{cond-mat/0303085}.

\bibitem[{\citenamefont{Moeller et~al.}(1995))\citenamefont{Moeller, Si,
  Kotliar, Rozenberg, and Fisher}}]{MSK+95}
\bibinfo{author}{\bibfnamefont{G.}~\bibnamefont{Moeller}},
  \bibinfo{author}{\bibfnamefont{Q.}~\bibnamefont{Si}},
  \bibinfo{author}{\bibfnamefont{G.}~\bibnamefont{Kotliar}},
  \bibinfo{author}{\bibfnamefont{M.}~\bibnamefont{Rozenberg}},
  \bibnamefont{and} \bibinfo{author}{\bibfnamefont{D.~S.}
  \bibnamefont{Fisher}}, \bibinfo{journal}{Phys. Rev. Lett.}
  \textbf{\bibinfo{volume}{74}}, \bibinfo{pages}{2082}
  (\bibinfo{year}{1995}).

\bibitem[{\citenamefont{Bulla}(1999)}]{Bul99}
\bibinfo{author}{\bibfnamefont{R.}~\bibnamefont{Bulla}},
  \bibinfo{journal}{Phys. Rev. Lett.} \textbf{\bibinfo{volume}{83}},
  \bibinfo{pages}{136} (\bibinfo{year}{1999}).

\bibitem[{\citenamefont{Bulla et~al.}(2001))\citenamefont{Bulla, Costi, and
  Vollhardt}}]{BCV01}
\bibinfo{author}{\bibfnamefont{R.}~\bibnamefont{Bulla}},
  \bibinfo{author}{\bibfnamefont{T.~A.} \bibnamefont{Costi}}, \bibnamefont{and}
  \bibinfo{author}{\bibfnamefont{D.}~\bibnamefont{Vollhardt}},
  \bibinfo{journal}{Phys. Rev. B} \textbf{\bibinfo{volume}{64}},
  \bibinfo{pages}{045103} (\bibinfo{year}{2001}).

\bibitem[{\citenamefont{Hirsch and Fye}(1986)}]{HF86}
\bibinfo{author}{\bibfnamefont{J.~E.} \bibnamefont{Hirsch}} \bibnamefont{and}
  \bibinfo{author}{\bibfnamefont{R.~M.} \bibnamefont{Fye}},
  \bibinfo{journal}{Phys. Rev. Lett.} \textbf{\bibinfo{volume}{56}},
  \bibinfo{pages}{2521} (\bibinfo{year}{1986}).

\bibitem[{\citenamefont{Schlipf et~al.}(1999))\citenamefont{Schlipf, Jarrell,
  van Dongen, Bl\"umer, Kehrein, Pruschke, and Vollhardt}}]{SJvD+99}
\bibinfo{author}{\bibfnamefont{J.}~\bibnamefont{Schlipf}},
  \bibinfo{author}{\bibfnamefont{M.}~\bibnamefont{Jarrell}},
  \bibinfo{author}{\bibfnamefont{P.~G.~J.} \bibnamefont{van Dongen}},
  \bibinfo{author}{\bibfnamefont{N.}~\bibnamefont{Bl\"umer}},
  \bibinfo{author}{\bibfnamefont{S.}~\bibnamefont{Kehrein}},
  \bibinfo{author}{\bibfnamefont{T.}~\bibnamefont{Pruschke}}, \bibnamefont{and}
  \bibinfo{author}{\bibfnamefont{D.}~\bibnamefont{Vollhardt}},
  \bibinfo{journal}{Phys. Rev. Lett.} \textbf{\bibinfo{volume}{82}},
  \bibinfo{pages}{4890} (\bibinfo{year}{1999}).

\bibitem[{\citenamefont{Rozenberg et~al.}(1999))\citenamefont{Rozenberg,
  Chitra, and Kotliar}}]{RCK99}
\bibinfo{author}{\bibfnamefont{M.~J.} \bibnamefont{Rozenberg}},
  \bibinfo{author}{\bibfnamefont{R.}~\bibnamefont{Chitra}}, \bibnamefont{and}
  \bibinfo{author}{\bibfnamefont{G.}~\bibnamefont{Kotliar}},
  \bibinfo{journal}{Phys. Rev. Lett.} \textbf{\bibinfo{volume}{83}},
  \bibinfo{pages}{3498} (\bibinfo{year}{1999}).

\bibitem[{\citenamefont{Joo and Oudovenko}(2001)}]{JO01}
\bibinfo{author}{\bibfnamefont{J.}~\bibnamefont{Joo}} \bibnamefont{and}
  \bibinfo{author}{\bibfnamefont{V.}~\bibnamefont{Oudovenko}},
  \bibinfo{journal}{Phys. Rev. B} \textbf{\bibinfo{volume}{64}},
  \bibinfo{pages}{193102} (\bibinfo{year}{2001}).

\bibitem[{\citenamefont{Bl\"umer}(2002)}]{Blu02}
\bibinfo{author}{\bibfnamefont{N.}~\bibnamefont{Bl\"umer}},
  \bibinfo{journal}{PhD thesis, Universit\"at Augsburg, 2002}.

\bibitem[{\citenamefont{Potthoff}(2003)}]{Pot03a}
\bibinfo{author}{\bibfnamefont{M.}~\bibnamefont{Potthoff}},
  \bibinfo{journal}{Euro. Phys. J. B} \textbf{\bibinfo{volume}{32}},
  \bibinfo{pages}{429} (\bibinfo{year}{2003}).

\bibitem[{\citenamefont{Baym and Kadanoff}(1961)}]{BK61}
\bibinfo{author}{\bibfnamefont{G.}~\bibnamefont{Baym}} \bibnamefont{and}
  \bibinfo{author}{\bibfnamefont{L.~P.} \bibnamefont{Kadanoff}},
  \bibinfo{journal}{Phys. Rev.} \textbf{\bibinfo{volume}{124}},
  \bibinfo{pages}{287} (\bibinfo{year}{1961}).

\bibitem[{\citenamefont{Potthoff et~al.}(2003)\citenamefont{Potthoff, Aichhorn,
  and Dahnken}}]{PAD03}
\bibinfo{author}{\bibfnamefont{M.}~\bibnamefont{Potthoff}},
  \bibinfo{author}{\bibfnamefont{M.}~\bibnamefont{Aichhorn}}, \bibnamefont{and}
  \bibinfo{author}{\bibfnamefont{C.}~\bibnamefont{Dahnken}},
  \bibinfo{journal}{Phys. Rev. Lett.} (in press).

\bibitem[{\citenamefont{Dahnken et~al.}())\citenamefont{Dahnken, Aichhorn,
  Hanke, Arrigoni, and Potthoff}}]{DAH+03}
\bibinfo{author}{\bibfnamefont{C.}~\bibnamefont{Dahnken}},
  \bibinfo{author}{\bibfnamefont{M.}~\bibnamefont{Aichhorn}},
  \bibinfo{author}{\bibfnamefont{W.}~\bibnamefont{Hanke}},
  \bibinfo{author}{\bibfnamefont{E.}~\bibnamefont{Arrigoni}}, \bibnamefont{and}
  \bibinfo{author}{\bibfnamefont{M.}~\bibnamefont{Potthoff}},
  \bibinfo{journal}{cond-mat/0309407}.

\bibitem[{\citenamefont{Gros and Valenti}(1993)}]{GV93}
\bibinfo{author}{\bibfnamefont{C.}~\bibnamefont{Gros}} \bibnamefont{and}
  \bibinfo{author}{\bibfnamefont{R.}~\bibnamefont{Valenti}},
  \bibinfo{journal}{Phys. Rev. B} \textbf{\bibinfo{volume}{48}},
  \bibinfo{pages}{418} (\bibinfo{year}{1993}).

\bibitem[{\citenamefont{S\'en\'echal et~al.}(2000)\citenamefont{S\'en\'echal,
  P\'erez, and Pioro-Ladri\`ere}}]{SPPL00}
\bibinfo{author}{\bibfnamefont{D.}~\bibnamefont{S\'en\'echal}},
  \bibinfo{author}{\bibfnamefont{D.}~\bibnamefont{P\'erez}}, \bibnamefont{and}
  \bibinfo{author}{\bibfnamefont{M.}~\bibnamefont{Pioro-Ladri\`ere}},
  \bibinfo{journal}{Phys. Rev. Lett.} \textbf{\bibinfo{volume}{84}},
  \bibinfo{pages}{522} (\bibinfo{year}{2000}).

\bibitem[{\citenamefont{Zacher et~al.}(2002))\citenamefont{Zacher, Eder,
  Arrigoni, and Hanke}}]{ZEAH02}
\bibinfo{author}{\bibfnamefont{M.~G.} \bibnamefont{Zacher}},
  \bibinfo{author}{\bibfnamefont{R.}~\bibnamefont{Eder}},
  \bibinfo{author}{\bibfnamefont{E.}~\bibnamefont{Arrigoni}}, \bibnamefont{and}
  \bibinfo{author}{\bibfnamefont{W.}~\bibnamefont{Hanke}},
  \bibinfo{journal}{Phys. Rev. B} \textbf{\bibinfo{volume}{65}},
  \bibinfo{pages}{045109} (\bibinfo{year}{2002}).

\bibitem[{\citenamefont{Kotliar et~al.}(2001))\citenamefont{Kotliar, Savrasov,
  P\'alsson, and Biroli}}]{KSPB01}
\bibinfo{author}{\bibfnamefont{G.}~\bibnamefont{Kotliar}},
  \bibinfo{author}{\bibfnamefont{S.~Y.} \bibnamefont{Savrasov}},
  \bibinfo{author}{\bibfnamefont{G.}~\bibnamefont{P\'alsson}},
  \bibnamefont{and} \bibinfo{author}{\bibfnamefont{G.}~\bibnamefont{Biroli}},
  \bibinfo{journal}{Phys. Rev. Lett.} \textbf{\bibinfo{volume}{87}},
  \bibinfo{pages}{186401} (\bibinfo{year}{2001}).

\bibitem[{\citenamefont{Bulla and Potthoff}(2000)}]{BP00}
\bibinfo{author}{\bibfnamefont{R.}~\bibnamefont{Bulla}} \bibnamefont{and}
  \bibinfo{author}{\bibfnamefont{M.}~\bibnamefont{Potthoff}},
  \bibinfo{journal}{Euro. Phys. J. B} \textbf{\bibinfo{volume}{13}},
  \bibinfo{pages}{257} (\bibinfo{year}{2000}).

\bibitem[{\citenamefont{Potthoff and Nolting}(1999)}]{PN99d}
\bibinfo{author}{\bibfnamefont{M.}~\bibnamefont{Potthoff}} \bibnamefont{and}
  \bibinfo{author}{\bibfnamefont{W.}~\bibnamefont{Nolting}},
  \bibinfo{journal}{Phys. Rev. B} \textbf{\bibinfo{volume}{60}},
  \bibinfo{pages}{7834} (\bibinfo{year}{1999}).

\bibitem[{\citenamefont{\=Ono et~al.}(2001){\natexlab{a}})\citenamefont{\=Ono,
  Bulla, and Hewson}}]{OBH01}
\bibinfo{author}{\bibfnamefont{Y.}~\bibnamefont{\=Ono}},
  \bibinfo{author}{\bibfnamefont{R.}~\bibnamefont{Bulla}}, \bibnamefont{and}
  \bibinfo{author}{\bibfnamefont{A.~C.} \bibnamefont{Hewson}},
  \bibinfo{journal}{Euro. Phys. J. B} \textbf{\bibinfo{volume}{19}},
  \bibinfo{pages}{375} (\bibinfo{year}{2001}{\natexlab{a}}).

\bibitem[{\citenamefont{\=Ono et~al.}(2001){\natexlab{b}})\citenamefont{\=Ono,
  Bulla, Hewson, and Potthoff}}]{OBHP01}
\bibinfo{author}{\bibfnamefont{Y.}~\bibnamefont{\=Ono}},
  \bibinfo{author}{\bibfnamefont{R.}~\bibnamefont{Bulla}},
  \bibinfo{author}{\bibfnamefont{A.~C.} \bibnamefont{Hewson}},
  \bibnamefont{and} \bibinfo{author}{\bibfnamefont{M.}~\bibnamefont{Potthoff}},
  \bibinfo{journal}{Euro. Phys. J. B} \textbf{\bibinfo{volume}{22}},
  \bibinfo{pages}{283} (\bibinfo{year}{2001}{\natexlab{b}}).

\bibitem[{\citenamefont{\=Ono et~al.}(2003))\citenamefont{\=Ono, Potthoff, and
  Bulla}}]{OPB03}
\bibinfo{author}{\bibfnamefont{Y.}~\bibnamefont{\=Ono}},
  \bibinfo{author}{\bibfnamefont{M.}~\bibnamefont{Potthoff}}, \bibnamefont{and}
  \bibinfo{author}{\bibfnamefont{R.}~\bibnamefont{Bulla}},
  \bibinfo{journal}{Phys. Rev. B} \textbf{\bibinfo{volume}{67}},
  \bibinfo{pages}{035119} (\bibinfo{year}{2003}).

\bibitem[{\citenamefont{Potthoff}(2001)}]{Pot01}
\bibinfo{author}{\bibfnamefont{M.}~\bibnamefont{Potthoff}},
  \bibinfo{journal}{Phys. Rev. B} \textbf{\bibinfo{volume}{64}},
  \bibinfo{pages}{165114} (\bibinfo{year}{2001}).

\bibitem[{\citenamefont{Runge and Gross}((1984))}]{RG84}
\bibinfo{author}{\bibfnamefont{E.}~\bibnamefont{Runge}} \bibnamefont{and}
  \bibinfo{author}{\bibfnamefont{E.~K.~U.} \bibnamefont{Gross}},
  \bibinfo{journal}{Phys. Rev. Lett.} \textbf{\bibinfo{volume}{52}},
  \bibinfo{pages}{997} (\bibinfo{year}{1984}).

\bibitem[{\citenamefont{Chitra and Kotliar}((2001))}]{CK01}
\bibinfo{author}{\bibfnamefont{R.}~\bibnamefont{Chitra}} \bibnamefont{and}
  \bibinfo{author}{\bibfnamefont{G.}~\bibnamefont{Kotliar}},
  \bibinfo{journal}{Phys. Rev. B} \textbf{\bibinfo{volume}{63}},
  \bibinfo{pages}{115110} (\bibinfo{year}{2001}).

\bibitem[{\citenamefont{Lin and Gubernatis}(1993)}]{LG93}
\bibinfo{author}{\bibfnamefont{H.~Q.} \bibnamefont{Lin}} \bibnamefont{and}
  \bibinfo{author}{\bibfnamefont{J.~E.} \bibnamefont{Gubernatis}},
  \bibinfo{journal}{Comput. Phys.} \textbf{\bibinfo{volume}{7}},
  \bibinfo{pages}{400} (\bibinfo{year}{1993}).

\bibitem[{\citenamefont{Kehrein}(1998)}]{Keh98}
\bibinfo{author}{\bibfnamefont{S.}~\bibnamefont{Kehrein}},
  \bibinfo{journal}{Phys. Rev. Lett.} \textbf{\bibinfo{volume}{81}},
  \bibinfo{pages}{3912} (\bibinfo{year}{1998}).

\bibitem[{\citenamefont{Noack and Gebhard}(1999)}]{NG99}
\bibinfo{author}{\bibfnamefont{R.~M.} \bibnamefont{Noack}} \bibnamefont{and}
  \bibinfo{author}{\bibfnamefont{F.}~\bibnamefont{Gebhard}},
  \bibinfo{journal}{Phys. Rev. Lett.} \textbf{\bibinfo{volume}{82}},
  \bibinfo{pages}{1915} (\bibinfo{year}{1999}).

\bibitem[{\citenamefont{Kotliar}(1999)}]{Kot99}
\bibinfo{author}{\bibfnamefont{G.}~\bibnamefont{Kotliar}},
  \bibinfo{journal}{Euro. Phys. J. B} \textbf{\bibinfo{volume}{11}},
  \bibinfo{pages}{27} (\bibinfo{year}{1999}).

\bibitem[{\citenamefont{Brinkman and Rice}(1970)}]{BR70}
\bibinfo{author}{\bibfnamefont{W.~F.} \bibnamefont{Brinkman}} \bibnamefont{and}
  \bibinfo{author}{\bibfnamefont{T.~M.} \bibnamefont{Rice}},
  \bibinfo{journal}{Phys. Rev. B} \textbf{\bibinfo{volume}{2}},
  \bibinfo{pages}{4302} (\bibinfo{year}{1970}).

\bibitem[{\citenamefont{Schiller and Ingersent}(1995)}]{SI95}
\bibinfo{author}{\bibfnamefont{A.}~\bibnamefont{Schiller}} \bibnamefont{and}
  \bibinfo{author}{\bibfnamefont{K.}~\bibnamefont{Ingersent}},
  \bibinfo{journal}{Phys. Rev. Lett.} \textbf{\bibinfo{volume}{75}},
  \bibinfo{pages}{113} (\bibinfo{year}{1995}).

\bibitem[{\citenamefont{Hettler et~al.}(1998))\citenamefont{Hettler,
  Tahvildar-Zadeh, Jarrell, Pruschke, and Krishnamurthy}}]{HTZ+98}
\bibinfo{author}{\bibfnamefont{M.~H.} \bibnamefont{Hettler}},
  \bibinfo{author}{\bibfnamefont{A.~N.} \bibnamefont{Tahvildar-Zadeh}},
  \bibinfo{author}{\bibfnamefont{M.}~\bibnamefont{Jarrell}},
  \bibinfo{author}{\bibfnamefont{T.}~\bibnamefont{Pruschke}}, \bibnamefont{and}
  \bibinfo{author}{\bibfnamefont{H.~R.} \bibnamefont{Krishnamurthy}},
  \bibinfo{journal}{Phys. Rev. B} \textbf{\bibinfo{volume}{58}},
  \bibinfo{pages}{R7475} (\bibinfo{year}{1998}).

\bibitem[{\citenamefont{Lichtenstein and Katsnelson}(2000)}]{LK00}
\bibinfo{author}{\bibfnamefont{A.~I.} \bibnamefont{Lichtenstein}}
  \bibnamefont{and} \bibinfo{author}{\bibfnamefont{M.~I.}
  \bibnamefont{Katsnelson}}, \bibinfo{journal}{Phys. Rev. B}
  \textbf{\bibinfo{volume}{62}}, \bibinfo{pages}{R9283}
  (\bibinfo{year}{2000}).

\bibitem[{\citenamefont{Maier et~al.}(2000))\citenamefont{Maier, Jarrell,
  Pruschke, and Keller}}]{MJPK00}
\bibinfo{author}{\bibfnamefont{T.}~\bibnamefont{Maier}},
  \bibinfo{author}{\bibfnamefont{M.}~\bibnamefont{Jarrell}},
  \bibinfo{author}{\bibfnamefont{T.}~\bibnamefont{Pruschke}}, \bibnamefont{and}
  \bibinfo{author}{\bibfnamefont{J.}~\bibnamefont{Keller}},
  \bibinfo{journal}{Euro. Phys. J. B} \textbf{\bibinfo{volume}{13}},
  \bibinfo{pages}{613} (\bibinfo{year}{2000}).

\end{thebibliography}
\end{document}